\documentclass{article}

\usepackage{PRIMEarxiv}

\usepackage[utf8]{inputenc} 
\usepackage[T1]{fontenc}    
\usepackage{hyperref}       
\usepackage{url}            
\usepackage{booktabs}       
\usepackage{amsfonts}       
\usepackage{nicefrac}       
\usepackage{microtype}      
\usepackage{lipsum}
\usepackage{fancyhdr}       
\usepackage{graphicx}       
\graphicspath{{media/}}     

\usepackage{cite} 
\usepackage{url} 
\usepackage[utf8]{inputenc} 
\usepackage{booktabs} 
\usepackage{graphicx}
\usepackage{lipsum}  
\usepackage{xcolor}
\usepackage{caption}

\usepackage[T1]{fontenc}
\usepackage{ragged2e}                       
\usepackage{booktabs, 
            makecell, multirow, tabularx}   

\title{Deep Fake Detection, Deterrence and Response: Challenges and Opportunities}

\author{
Amin Azmoodeh, Ali Dehghantanha\\
		Department of Computer Science\\
		University of Guelph, ON, Canada\\
		aazmoode@uoguelph.ca ; adehghan@uoguelph.ca
}

\begin{document}
\maketitle
\begin{abstract}
According to the 2020 cyber threat defense report, 78\% of Canadian organizations have experienced at least one successful cyberattack in 2020 \cite{web1}. The consequences of such attacks vary from privacy compromises to immerse damage costs for individuals, companies, and countries. Specialists predict that the global loss from cybercrime will reach 10.5 trillion US dollars annually by 2025 \cite{web2}. Given such alarming statistics, the need to prevent and predict cyberattacks is as high as ever. The explosive evolution of AI as cutting-edge contemporary technology is changing the face of the globe as it is penetrating almost all aspects of our life. AI has brought us agile, accurate, and cost-effective solutions for sweeping domains, including but not limited to medicine, finance, engineering, and cybersecurity. Our critical infrastructure is increasingly relying on Artificial Intelligence (AI) and Machine Learning (ML)-based solutions to offer timely services at scale \cite{8476541}. Our increasing reliance on ML-based systems raises serious concerns about the security and safety of these systems. Especially the emergence of powerful machine learning techniques to generate fake visual, textual, or audio content with a high potential to deceive humans raised serious ethical concerns. These artificially crafted deceiving videos, images, audio, or texts are known as Deepfakes garnered attention for their potential use in creating fake news, hoaxes, revenge porn, and financial fraud. Even more recently in the Russia-Ukraine conflict, deepfakes are used to change the course of war\cite{web4}. Diversity and the widespread of deepfakes made their timely detection a significant challenge. In this paper, we first offer background information and a review of previous works on the detection and deterrence of deepfakes. Afterward, we offer a solution that is capable of 1) making our AI systems robust against deepfakes during development and deployment phases; 2) detecting video, image, audio, and textual deepfakes; 3) identifying deepfakes that bypass detection (deepfake hunting); 4) leveraging available intelligence for timely identification of deepfake campaigns launched by state-sponsored hacking teams; 5) conducting in-depth forensic analysis of identified deepfake payloads. Our proposed solution can be used as a technical guide for developing detection, deterrence, and forensics investigation solutions for deepfakes. Our solution would address important elements of Canada’s National Cyber Security Action Plan (2019-2024) in increasing the trustworthiness of our critical services \cite{web5}. Following actions can be taken based on this research findings: 

\begin{itemize}
		\item \textbf{Raising public awareness about risks of deepfakes:} increasing the understanding of deepfake threats and empowering Canadian public to do their part in keeping our society and critical services safe from deepfake-based attacks is the most important and effective step in reducing risk of deepfakes. Cybersecurity should always be considered as a shared responsibility. While this paper is focused on development of technical solutions for early detection and deterrence of deepfakes,  the effectiveness of our solutions (or any technical solution in cybersecurity) are limited without regular and systemic public awareness campaigns. Supporting development of public training programs in this domain should be considered as a top priority.  
		\item \textbf{Developing AI robustness monitoring solutions:} there is a growing trend in using AI to detect deepfakes. However, more recently, adversaries made attempts to create adversarial deepfake payloads that are capable of deceiving humans while bypassing AI-based detection systems! Susceptibility of AI systems to adversarial payloads made them vulnerable to deepfake adversarial contents. Hence, any AI-based deepfake detection system should be monitored during development and after deployment to assure its robustness against adversarial payloads.
		\item \textbf{Addressing the risk of biasness in AI-based deepfake detection and deterrence systems:} Our detailed literature review demonstrated that existing deepfake detectors are negatively biased toward specific  features of payloads. For instance, focusing on eye blinks, exploring specific unusual features from a speech, or inconsistency in a text. However, this biasness could lead to unintentional filtering of benign contents generated by marginal groups. Hence, it is important to always monitor and evaluate the biasness risk of these systems. Our recommendation is to consider building multi-view detection/deterrence systems that consist of several local mechanisms each of which validates different micro-feature of payloads. Such system could also consider macro-features at a higher level. For instance, it can decide based on the inconsistency among speech content, picture, and voice of a public speaker all at the same time. 
		\item \textbf{Assuring explainability of deepfake detection and deterrence systems:} Explainability of the decisions made by the deepfake detectors is a crucial requirement. This not only helps in proving the forensic soundness of the system but provides means to make sure that all system decisions are made with legal, ethical, and privacy considerations.
		\item \textbf{Development of offensive deepfake technologies:} finally, it is important to note that we observed a fast growth in using deepfakes by advanced state-sponsored hacking groups as part of both classical and cyber war campaigns. For example, deepfake content was heavily used as part of the recent Russia-Ukraine clash to impact public opinion, distribute false information, or even mislead troops on the ground. We believe offensive deepfake technologies could play an even more significant role on battlefields for deceiving both human and AI systems. For example, deepfake generation technologies can suggest fake troop formation or types of equipment on the battleground to deceive warlords' cognition or AI-based object detection systems. 

	\end{itemize}
\end{abstract}
\newpage
\tableofcontents
\listoffigures
\listoftables
\newpage



	\section{Introduction}\label{sec:intro}
\begin{figure*}[h]
			\includegraphics[width=\textwidth,height=10cm]{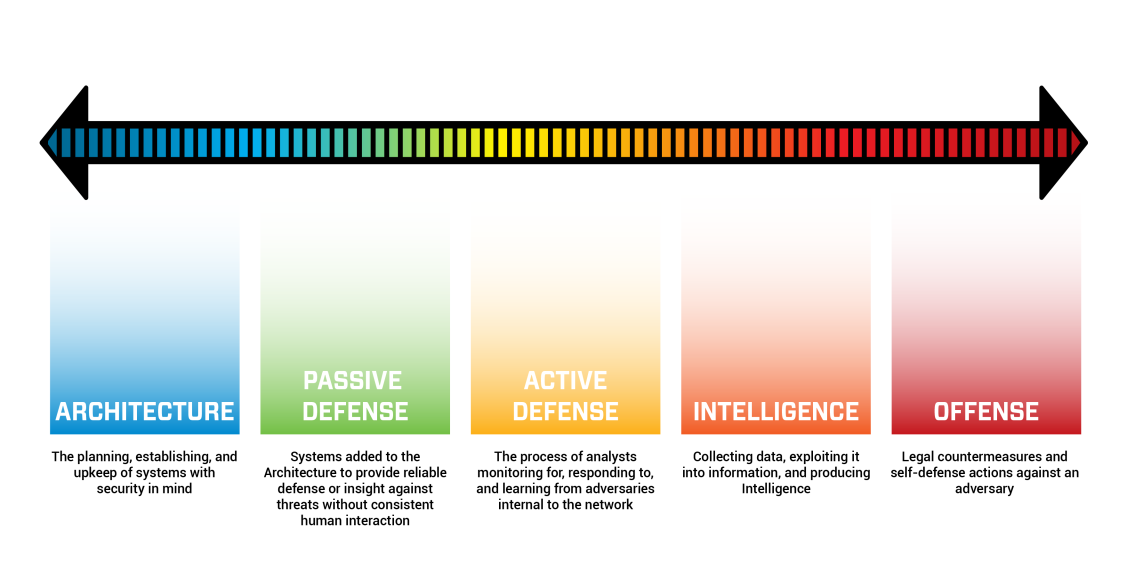}
			\caption{The Sliding Scale of Cybersecurity}
			\label{fig:slidingscale}
		\end{figure*}

\begin{figure*}[h]
			\includegraphics[width=\textwidth,height=7.5cm]{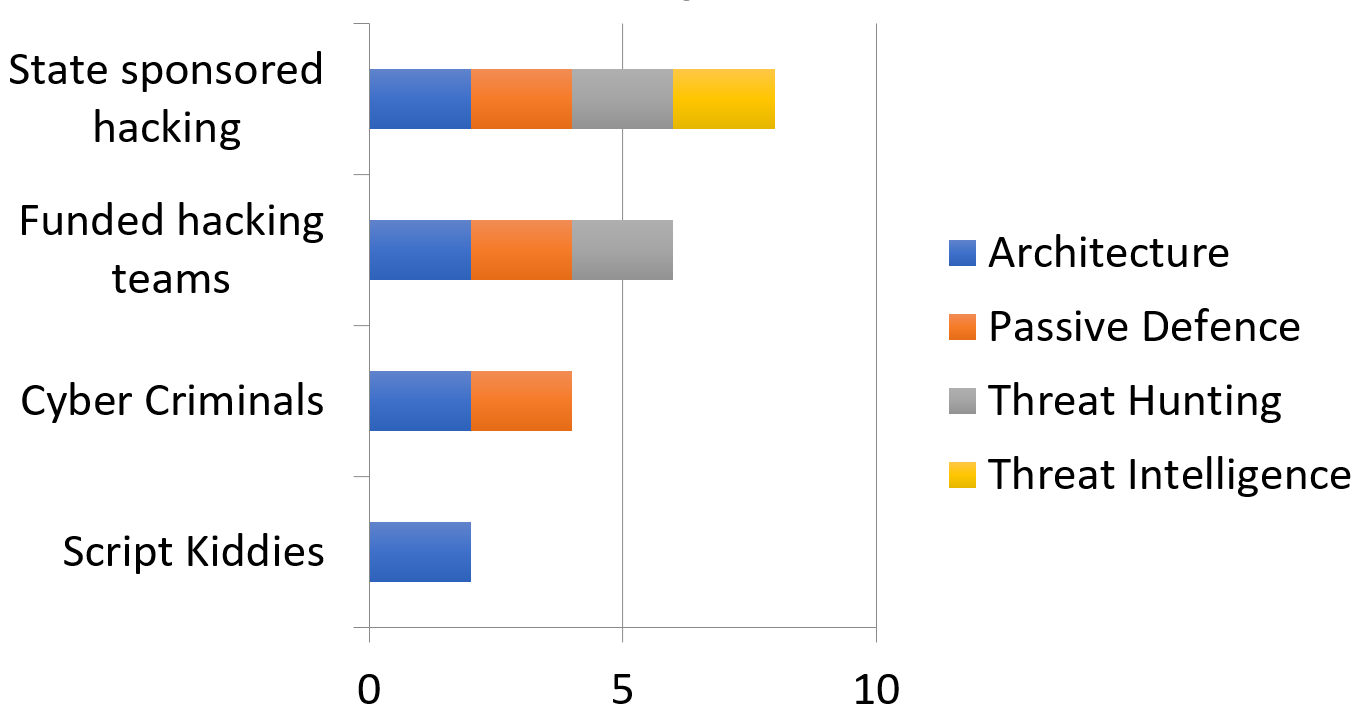}
			\caption{Addressing Risk of Adversaries}
			\label{fig:adversaryrisk}
		\end{figure*}

\begin{figure*}[h]
		\includegraphics[width=\textwidth,height=8cm]{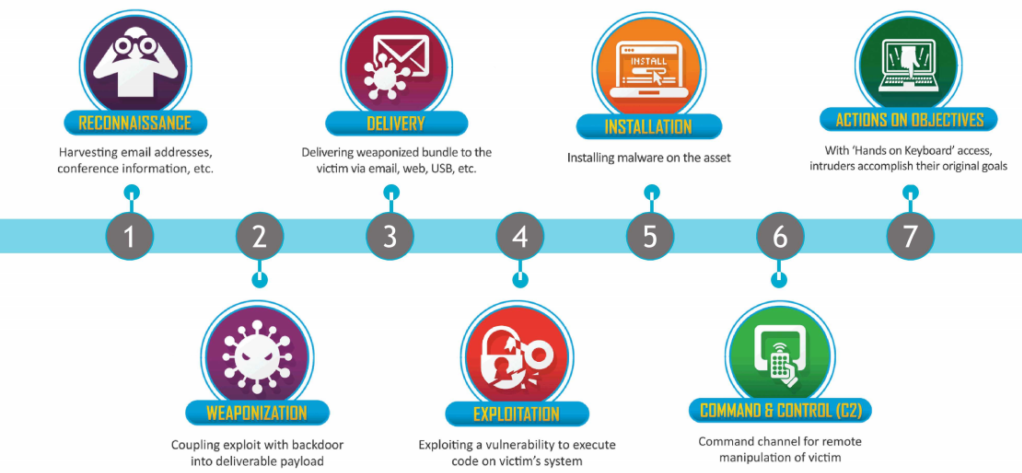}
		\caption{Cyber Kill Chain \cite{misc-cyberkillchain}}
		\label{fig:ckc}
	\end{figure*}

	For most, it would be hard to imagine a world without digitally connected systems managing almost all aspects of our day to day life. These ubiquitous services and applications are growing and impacting nearly everything from our social lives to today's bleeding edge businesses. However, the ubiquitousness of these technologies has also made them a prime driver of exasperated cybercrime. This heightened level of threats has created an arms race between thousands of cybersecurity businesses and their advanced adversaries. Security companies around the globe are designing new technologies to protect digital systems while on the other hand, their adversaries spend their time developing exploits designed to take advantage of vulnerabilities, flaws, and misconfigurations within today's complex cyber architectures. The arms race between the cyber defender and cyber attacker generated a multi billion-dollar industry that is predicted to grow year over year \cite{costello_2019}. This cyber arms created the fifth dimension of warfare (cyber warfare) which quickly becomes the central battlefield in many conflicts. Classical warfare was geographically bound, hence, it was possible to segregate and identify the enemy. However, the cyber warfare is not bound by geographic location, making it increasingly difficult to defend. Further, the lack of consequence blurs the lines between soldier and civilian in cyber warfare while it is the cheapest means of warfare, and as such an attractive opportunity for nation states. Private businesses are often leveraged for intelligence to deploy these cyber-attacks by nation states or more commonly referred to as Advanced Persistent Threats (APTs).\\
	The Sliding Scale of Cybersecurity (SSC) (see Figure \ref{fig:slidingscale}) is built as a model for categorizing actions and investments in cybersecurity. This model forms the basis for designing cybersecurity technologies that protect enterprise businesses. Different layers of the SSC model attempts to address different categories of adversaries (see Figure \ref{fig:adversaryrisk}). Secure architecture as the first and the most basic layer, addresses the risk of script kiddies. These adversaries have very limited skills and majority are using existing malicious payloads to target organizations. Authentication, authorization, identity management and access control systems are among technologies deployed in this layer. The second layer contains passive defense mechanisms.  Classic cybersecurity technologies such as those focused on file hashes and static signatures to identify malicious software (malware) are deployed in this layer. Current state-of-the-art systems may use a combination of conventional ML for more basic threats and deep learning (DL) systems to detect some of the more advanced threats in this layer. This improvement is sufficient for some of the more direct cyber threats but trivial to evade for APT's or funded hacking teams \cite{jcp1030021}. Active defense A.K.A Threat Hunting is the third layer of cybersecurity where defenders try to detect attackers bypassed passive and architecture-level security mechanisms. Funded hacking teams that have enough resources to bypass passive and architecture-level security mechanisms should be identified in this layer. The next layer is threat intelligence where we use known tactics, techniques and procedures (TTPs) of (mostly) APT actors to identify their activities before they meet their objectives.  At this layer we are mainly dealing with state-sponsored hacking teams. Due to the influx of APTs, businesses and countries have valued monitoring and understanding them further; This process is called threat intelligence or cyber threat intelligence (CTI). Threat intelligence is data that is collected, processed, and analyzed to understand a threat actor’s motives, targets, and attack behaviours. Threat intelligence enables defenders to make faster, more informed, data-backed security decisions and change their behavior from reactive to proactive in the fight against threat actors \cite{marketing-crowdstrike-001}. However valuable threat intelligence may be, it is not always enough to fully defend a network. Often threat intelligence is only helpful if applied correctly to the organization and require cyber threat hunters, to test, verify and validate existing security controls to maximize the use of threat intelligence. The formalization of cyber threat hunting and cyber threat intelligence over the past decade has proven helpful in many ways to the cybersecurity industry in regards to detecting APT activity. Cyber threat hunting is a focused and iterative approach to searching out, identifying, and understanding adversaries who have entered the defender's networks. A recent survey of 600 respondents shows most organizations still do not fully understand cyber threat hunting \cite{lee_lee_2018}. Although threat hunting is not a new practice, the formalization of the art and standard operating procedures are becoming more widely defined as APTs become more and more of a risk. Cyber threat hunting has increased the detection of advanced threats and reduced the dwell time of an APT within a network (Dwell time is defined as the time between an adversary archives initial access within a network until its presence is detected). Due to the significant time investment, reducing the dwell time, or mean time to detect (MTTD), is critical. Global information security leader FireEye started tracking the average dwell time publicly in their yearly M-Trends reports starting in 2011. Since then, we have seen dwell time reduced from 416 days to less than 30 days on average in North America as of 2020.Popular frameworks such as the Diamond Model of analysis \cite{misc-diamond-model}, the cyber kill chain \cite{misc-cyberkillchain} (see Figure \ref{fig:ckc}), and most recently, MITRE ATT\&CK \cite{misc-mitreattack} provide an outline for cyber threat hunters to assess the situation and further reduce the dwell time. When evaluating cyber threat hunting programs, there are many metrics to determine the maturity of the hunt. An effective program divides its metrics into tactical, operational, and strategic categories.  An example of a tactical metric is the efforts to map attack campaigns to models such as the MITRE ATT\&CK framework. Mapping threats to the standard model allows operators to deploy defences more effectively. Operational metrics could be the time it takes an analyst to conclude, the number of hunts generated from a data source, number of discovered hunt findings, or the number of gaps in data sources identified. Finally, an example of a strategic metric could be the number of successful hunts performed. Many other metrics exist to measure the effectiveness of cyber threat hunting but vary organization to organization based on their appetite for risk. Generally the goal is to detect undetected threats while identifying gaps within the security posture, and one approach to automate this is ML. Finally, the last layer of SSC includes forensics and investigative techniques to take legal and offensive actions against adversaries. 
	

\begin{figure}[ht]
		\centering
		\includegraphics[width=12cm]{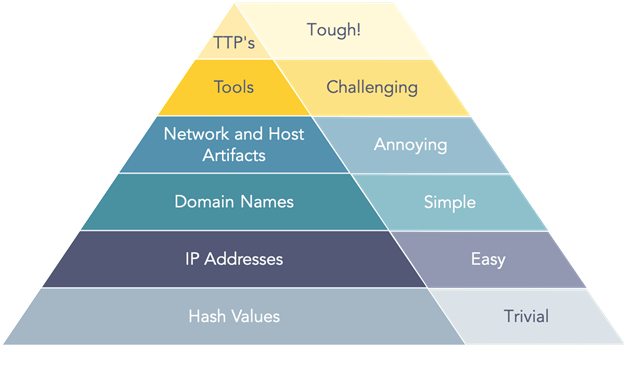}
		\caption{David Bianco's Pyramid of Pain \cite{bianco_2013}}
		\label{fig:PoP}		
	\end{figure}

\begin{figure}
	 	\centering
	 	\includegraphics[width=0.7\linewidth]{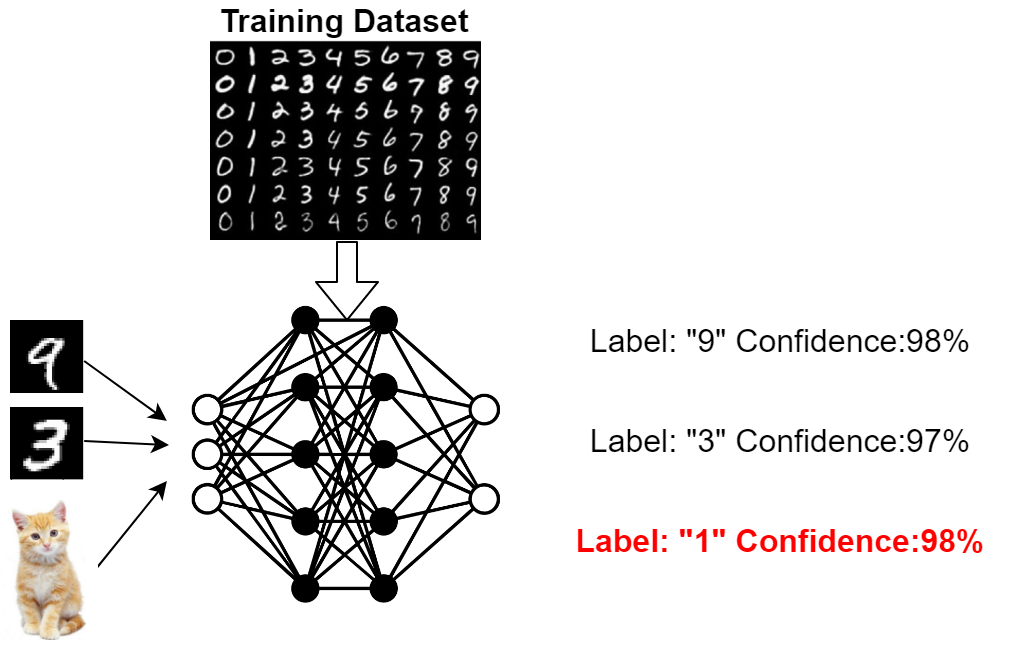}
	 	\caption{A typical DCNN model trained with MNIST dataset and generates confidence scores for Cat image as input}
	 	\label{fig:ood}
	 \end{figure}

\begin{figure*}[h]
		\includegraphics[width=\textwidth,height=11cm]{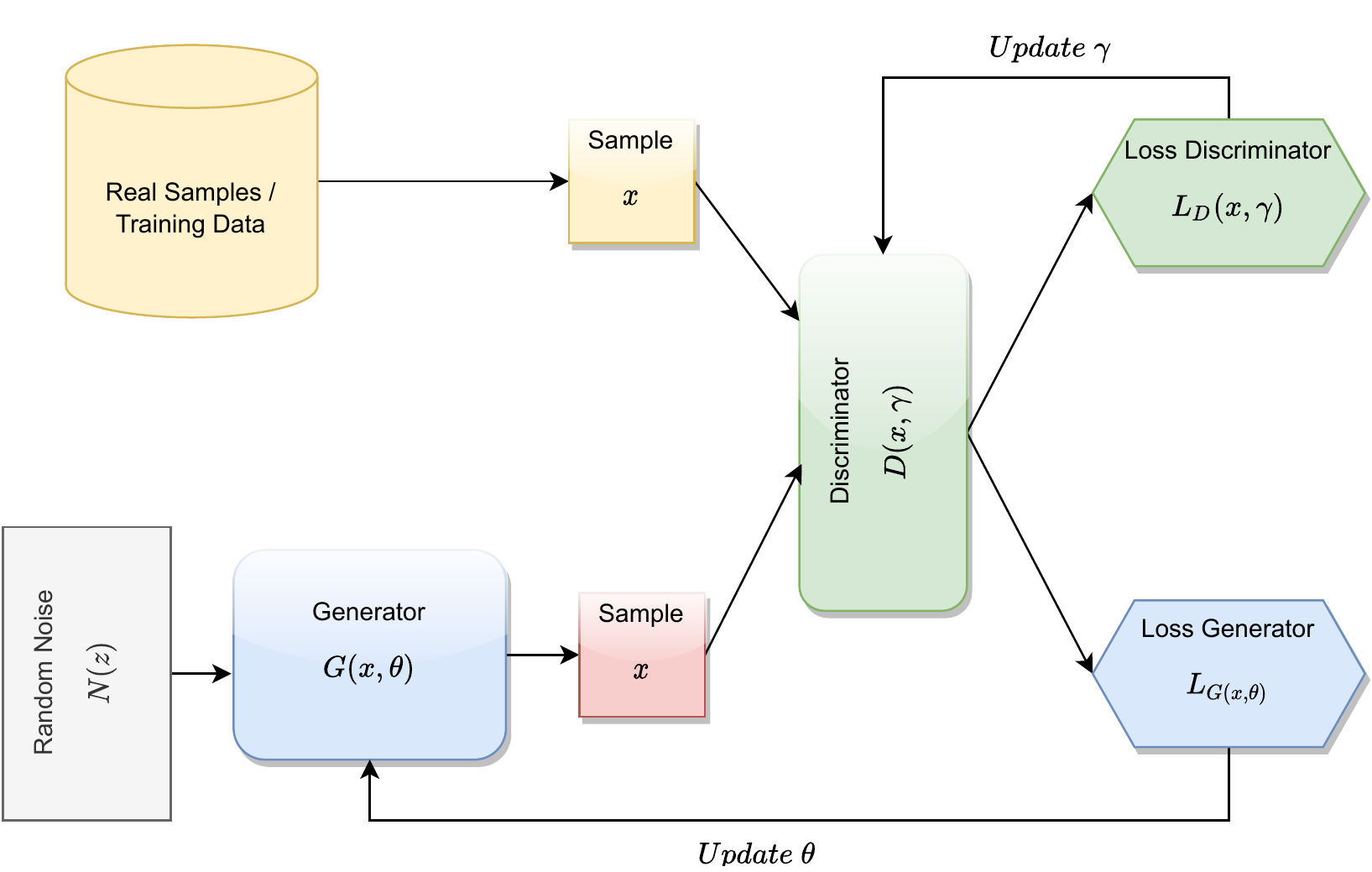}
		\caption{Architecture of a typical Generative Adversarial Network.}
		\label{fig:gan}
	\end{figure*}

	The pyramid of pain (see Figure \ref{fig:PoP}) \cite{bianco_2013}, is a commonly used diagram to describe the types of indicators of compromise (IoCs) and their relation to challenges for an adversary to change its behaviour. The right side legend describes the level of pain an adversary goes through for changing an IoC in different layers. 
	 As the quantity of data grew year over year, it would become more challenging to review and find threats manually. ML-enhanced intrusion detection and prevention systems offer far better anomaly detection that does not require manual security engineering efforts. However, the engineering effort was not wholly removed but displaced to machine learning tuning, feature engineering, and data analysis tasks.  DL solutions are possibly the most widely adopted subset of ML models that outperform classical algorithms and even human beings in a variety of tasks, including object recognition, malware detection, financial predictions, and fraud analysis\cite{liu2017survey}. DL networks perform automatic feature extraction by connecting and identifying complex patterns and embeddings capable of creating incredibly accurate models. These models would take considerable time and data to train. The next issue with DL is that it is tough to explain why a model converges the way it does, and many analysts found it hard to trust these mysterious embeddings when making critical decisions. However, the wide adoption of DL in  safety critical systems has raised concerns over the security and safety of DL solutions \cite{papernot2018sok}. In the last decade many researchers attempted challenges in various aspects of DL's safety, such as robustness\cite{pei2017deepxplore,guo2018dlfuzz,odena2019tensorfuzz}, fairness\cite{kusner2017counterfactual} and privacy\cite{bichsel2018dp,8835245}.\\
	 A typical DL model is expected to deal with input samples that are similar to its training data. However, in the real world, a trained DL may encounter inputs unlike anything that it has been trained with In-Distribution (ID) samples refer to a DL model input drawn or generated from the model's training data. In contrast, Out-of-distribution (OOD) refers to samples that are not comparable with ID samples. It is expected that a DL model offers a low confidence score to an OOD sample. However, it is not uncommon for a DL model to assign a high confidence score even to an OOD sample\cite{HendrycksG17} (See Fig. \ref{fig:ood}). These high confident OOD samples supposed to produce low confidence outputs while unexpectedly are resulting to a high confident outputs are comprising the attack surface of the given DL. \\
    Deepfake payloads (stemming from “deep learning” and “fake”) are originally created by techniques that generate synthetic contents that imitate human beings such as the human face, speech, or written texts. Deepfake payload generation techniques utilize DL algorithms to manipulate given inputs and generate authentic-looking fake payloads \cite{10.1145/3425780}.  These generative techniques to create semi-realistic contents quickly show their adversarial applications in exploiting human trust over digital contents \cite{zhang2019adversarial}. 
	Recently, significant advancement in DL models as well as computational techniques, encouraged researchers for inventing revolutionary DL models that generate deepfake payloads \cite{10.1145/3425780}. Although majority of generators are targeting imagery tasks including facial image and video manipulation\cite{tyagi2021comprehensive}, generating fake voices \cite{voicekhanjani2021deep} and fake news\cite{zhou2018fake,de2021survey} are other important trends. Advent of Generative Adversarial Networks (GAN), \cite{goodfellow2014generative,cai2021generative,gui2021review,wang2021generative,creswell2018generative, jabbar2021survey, pumarola2020ganimation, sanchez2018triple}, revolutionized technologies, and tools for synthetic content generation. For instance and in 2020, approximately 78 papers per day were published using GAN technology. GANs have been widely leveraged for fake payload generation including image and video \cite{wang2021generative}, news \cite{de2021survey} and voice \cite{latifi2019audio}.A typical GAN includes a generator ($G(x,\theta)$) and a discriminator component ($D(x,\gamma)$) that are technically - but not exclusively - implemented by DL models ($\theta$ and $\gamma$ represent internal parameters for the given input $x$ - see Figure \ref{fig:gan}). The generator objective is to understand the distribution of real samples to generate new fake payloads using a random noise $N_z$ function. On the other hand, the discriminator which is usually a binary classifier, is responsible for differentiating between real and fake samples. The optimization approach of GANs can be considered as a \textit{minimax} game. The optimization terminates at a saddle	point that forms a minimum with respect to the generator and a maximum with respect to the discriminator. Hence, the GAN optimization goal is to reach a Nash equilibrium between the generator and discriminator components \cite{gui2021review}. \\
    In this paper, we offer a comprehensive solution aligned with SSC for preventing, detecting, hunting and investigating Deepfake payloads. While practical, our solution covers all layers of SSC and aligns with current state of the art in the field.   \\

	\section{Deep Fake: The Uses and Abuses}\label{sec:useabuse}
In this section we explain main trends for generating deepfake payloads including fake image and video contents, fake voice data and fake news. While we review most significant related works in this section, an enthusiastic reader may refer to the curated list of surveys at the end of this section for further reading (see Table \ref{tab:generation}). 

	\subsection{DeepFake Images and Videos} 
\begin{figure*}[h]
		\includegraphics[width=\textwidth]{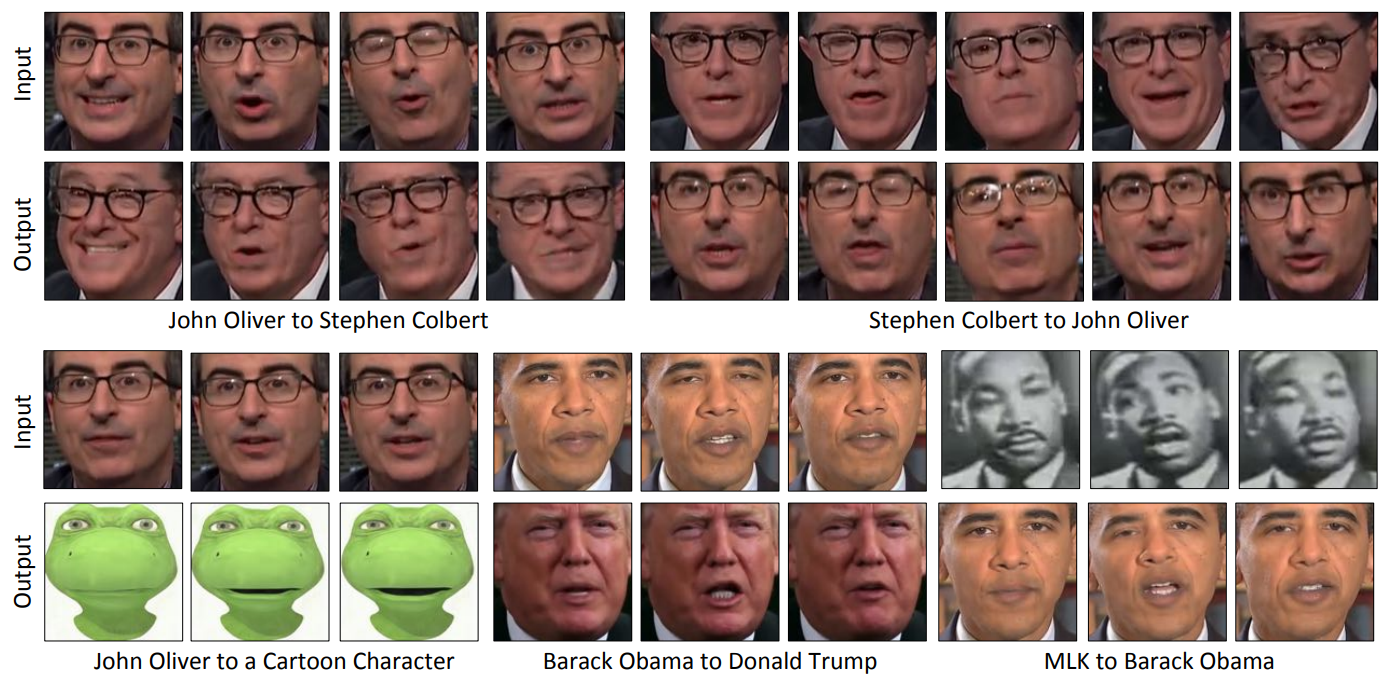}
		\caption{\textit{RecycleGAN} \cite{bansal2018recycle} input to output image translation for generating deepfake samples}
		\label{fig:RecycleGanSamples}
	\end{figure*}

\begin{figure*}[h]
		\includegraphics[width=\textwidth]{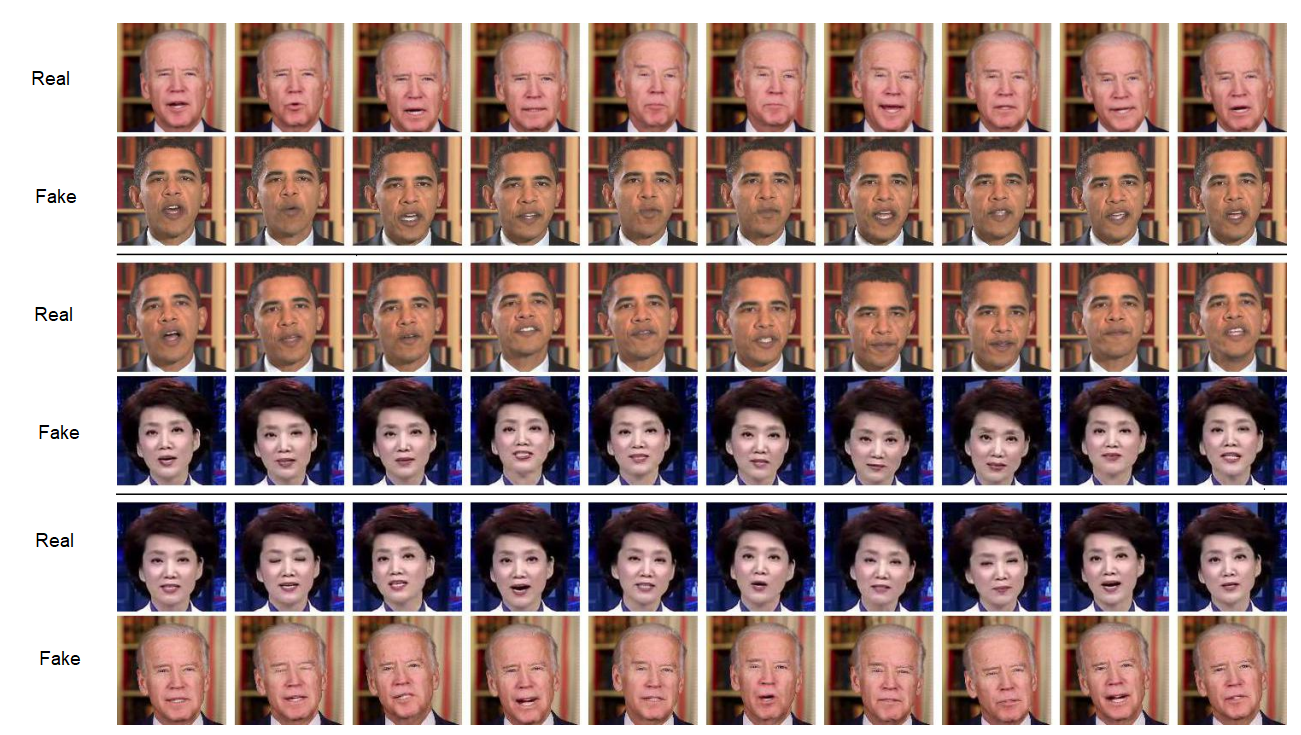}
		\caption{XU et al.\cite{xu2017face} input frames to output frames translation for generating deepfake videos}
		\label{fig:XUSamples}
	\end{figure*}

\begin{figure*}[h]
		\includegraphics[width=\textwidth]{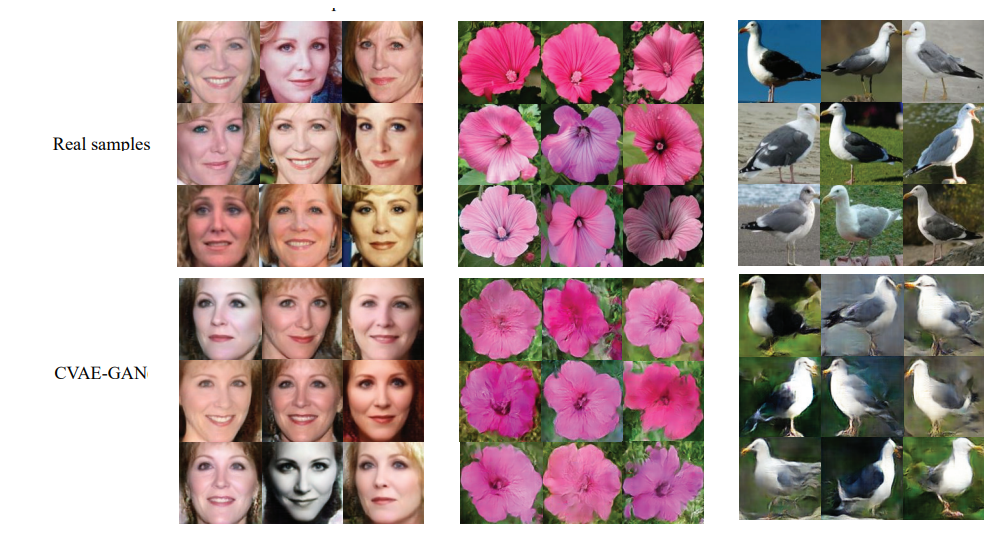}
		\caption{Bao et al.\cite{bao2017cvae} input and output generated by \textit{CVAE-GAN}}
		\label{fig:CVAEGANSamples}
	\end{figure*}

\begin{figure*}[h]
		\includegraphics[width=\textwidth]{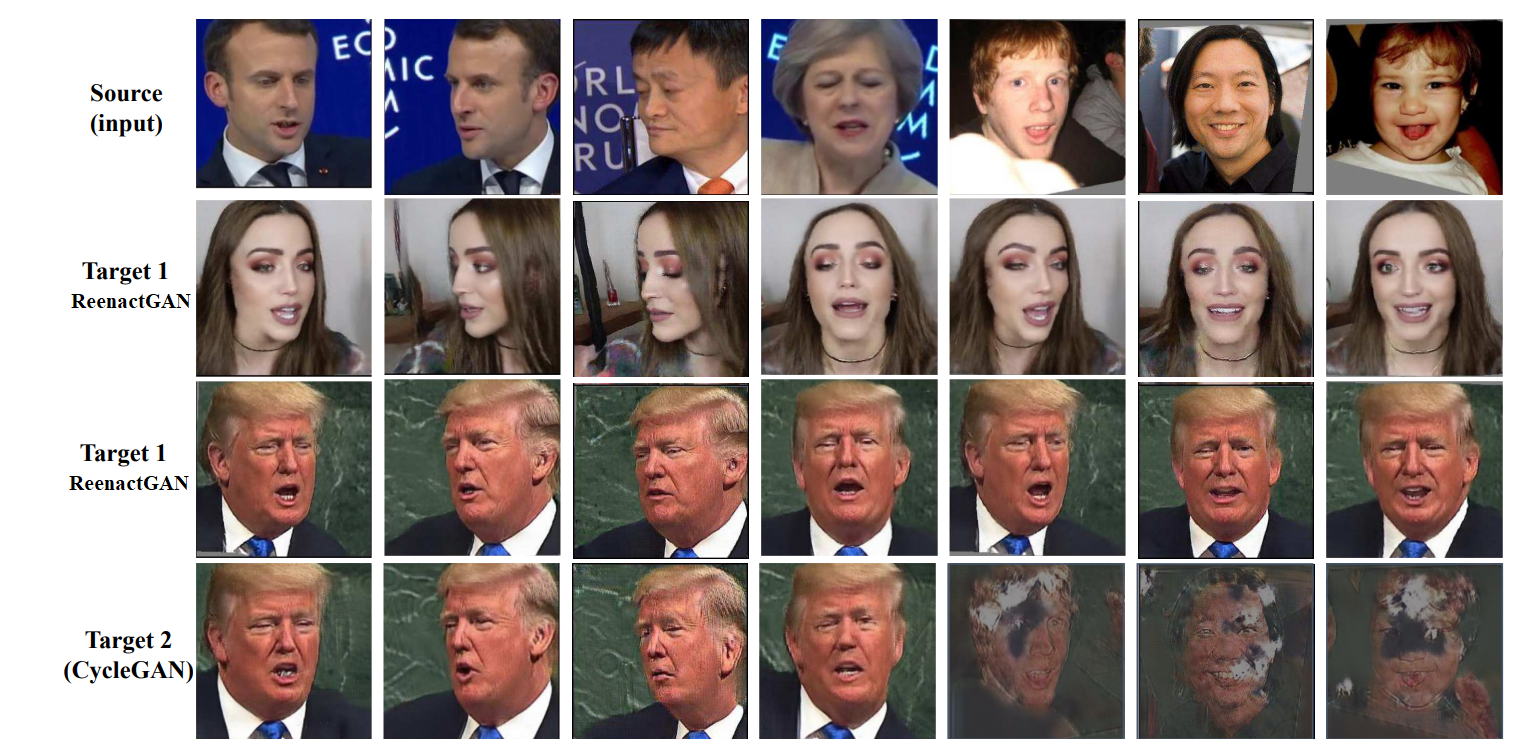}
		\caption{Wu et al.\cite{bao2017cvae} input and output generated by \textit{ReenactGAN} and comparison with \textit{CycleGAN} for large facial movements}
		\label{fig:ReenactGANSamples}
	\end{figure*}

\begin{figure*}[h]
		\includegraphics[width=\textwidth]{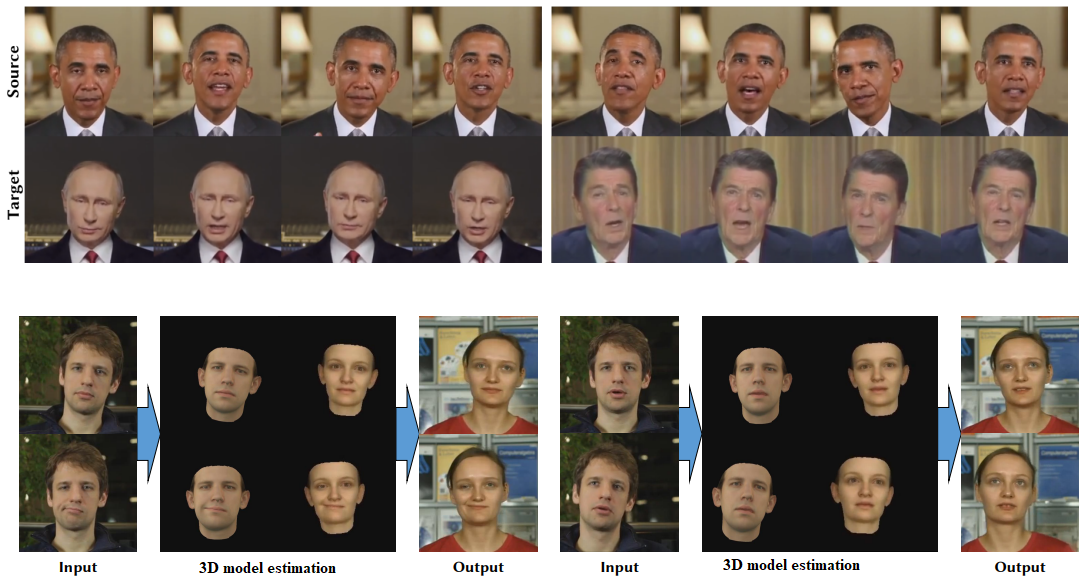}
		\caption{Kim et al.\cite{kim2018deep} input and output as well as 3D model estimation}
		\label{fig:KimSamples}
	\end{figure*}

 Fake images and videos are the most common payload in the deepfake domain. This could be due to their further appeal to human audience. A wide range of methods have been proposed to produce synthetic visionary contents, and GANs play a pivotal role in this task.  \\
	Using \textit{CycleGAN} \cite{Isola_2017_CVPR} which is a GAN for image to image translation, Bansal et al. \cite{bansal2018recycle} proposed a fake image and video generator and called \textit{RecycleGAN}\footnote{See generated video samples by RecycleGan here: \url{https://www.cs.cmu.edu/~aayushb/Recycle-GAN/} (Accessed:March 2022)}. Their proposed generator was a  generic translation network that enhanced temporal coherence by including next-frame predictor networks in each domain. For fake face generation, the proposed method's network was trained to translate the facial properties of $x_{source}$ into portraits of $x_{target}$. Figure \ref{fig:RecycleGanSamples} illustrates \textit{RecycleGAN} image translation and generated samples.\\
	Using a similar approach, Xu et al. \cite{xu2017face} proposed a deepfake video generator based on \textit{CycleGAN} and one-to-one video frame translation (see Figure \ref{fig:XUSamples}). Although, their method did not require data pairing, but it needed shared properties between source and domain inputs such as head pose and facial expression. \\
	Bao et al.\cite{bao2017cvae} proposed a \textit{many-to-one} generative architecture that combined Variational Auto Encoders (VAE) \cite{asperti2021survey} and GANs to generate many samples (identity) based on given input $x$ as shown in Figure \ref{fig:CVAEGANSamples}.  The VAE component generated multiple inputs for $G(x,\theta)$ that offered more stable training for $G(x,\theta)$ while keeping the structure of input $x$. \\
	In order to support large facial movements of the source compared with \textit{CycleGAN} and \textit{RecycleGAN}, Wu et al. \cite{wu2018reenactgan} presented \textit{ReenactGAN}. This method transferred the image boundary instead of its pixel space. Hence, the transferring process could be completed in almost real-time on a traditional workstation while keeping the photo-realistic face reenactment. Figure \ref{fig:ReenactGANSamples} gives a comparative  overview of \textit{ReenactGAN} and \textit{CycleGAN}  in presence of large movements.\\
	Tulyakov et al. \cite{tulyakov2018mocogan} proposed \textit{MoCoGAN} that considered motion and content of source objects using a Recurrent Neural Network (RNN) to improve image generation. The RNN component was responsible for monitoring past frames information to reduce the loss while the source object was moving. Therefore,  \textit{MoCoGAN} could generate realistic and coherent frames. Generated fake videos demonstrated significant improvements for moving objects such as athletic activities. Wang et al.\cite{wang2018video} introduced  \textit{Vid2Vid} as an advanced video synthesizing framework \footnote{See Vid2Vid \cite{wang2018video} samples here : \url{https://www.youtube.com/watch?v=GrP_aOSXt5U}(Accessed March 2022)}. The framework included the temporal aspects for generating $L+1$ frame using the source frames and any generated fake frames of 1 to $L$. \\
	For generating more realistic facial fake videos, Kim et al.\cite{kim2018deep} proposed a method to generate a 3D model of a face from given 2D input using monocular reconstruction and then transferred the source to the target based on the extracted 3D model as shown in Figure \ref{fig:KimSamples}. Their model considered several facial attributes, including pose, eye gaze, etc. Their experiments demonstrated significant improvements in genuine face imitation, especially in facial expressions.
	
	\subsection{DeepFake Voice}
		 Hearable deepfake payloads are probably the most widely misused deepfake content among criminals. Deepfake voice payloads can be used to bypass voice-based authentication systems or as part of social engineering attacks. Unlike image/video deepfakes that are mostly generated using GANs, deepfake speech and music generators are using a wide range of techniques. In this section, we review the most prominent techniques for generating deepfake voice contents. \\
		Zen et al.\cite{zen2007hmm} proposed a method using Hidden Markov Models (HMM) with an adaptive training method and a parameterized speech generation algorithm. HMM generates its parameters based on the textual representation of the given voice input which makes this model capable of generating deepfake voice payloads with different characteristics and in different languages. \\
	Oord et al. \cite{oord2016wavenet} introduced \textit{WaveNet} which was the first model that could use raw audio waveform signals instead of acoustic features of the voice. It had an auto-regressive and probabilistic core that could be trained with waveform signals. The \textit{WaveNet} outperformed parametric and statistical approaches in generating natural sounds that were indistinguishable even by a human listener. Many researchers improved the \textit{WaveNet} to produce even a more natural sound such as \textit{Fast WaveNet}\cite{paine2016fast}, \textit{Parallel WaveNet}\cite{oord2018parallel}, and \textit{Flow WaveNet}\cite{kim2018flowavenet}.\\
		\textit{Deep Voice} is another major family of speech generators first introduced by Arik et al.\cite{arik2017deep}. The \textit{Deep Voice}'s architecture includes four fundamental components to deliver its speech-to-text functionality. The first component is a segmented convolutional neural network for identifying boundaries of phonemes. The second component is a multi-layer decoder/encoder designed to convert graphemes to phonemes. The third component estimates phonemes length and their frequencies, and the fourth component is a \textit{WaveNet} network that generates the final synthetic voice payload. \textit{Deep Voice-2} \cite{gibiansky2017deep} aimed to  support multi-speaker and \textit{Deep Voice-3} \cite{ping2017deep} attempted to improve the quality of generated voice. Google developed \textit{Tacotron} \cite{wang2017tacotron} as a seq-to-seq model that uses an attention-based mechanism. At the core of \textit{Tacotron}, a module was responsible for extracting new representations from inputted voice sequence. Afterwards, a content-oriented attention model was utilized to feed extracted representations to a decoder. The decoder utilized a recurrent approach to generate a waveform sound file. \textit{Tacotron} was the basis for many future researches such as \cite{weiss2021wave,ling2013modeling,shen2018natural}. \\
		Binkowski et al.\cite{binkowski2019high} introduced \textit{GAN-TTS}, as a GAN-based approach to create high fidelity speech contents.  \textit{GAN-TTS} leveraged a conditional feed-forward generator to produce raw audio files. Afterwards, an ensemble of discriminative models were used to improve properties of generated audio files and make them more natural. \textit{GAN-TTS} achieved significant success in terms of human approval. Continuing this approach,  EATS\cite{donahue2020end} was proposed by Donahue et al. to increase the quality of generated waveform sound files. \\
		Kumar et al. \cite{kumar2019melgan} presented another types of GAN-based speech generators called \textit{MelGAN}. The presented generator was non-autoregressive, fully convolutional, with very few parameters. \textit{MelGAN} showed a superior performance in terms of generating quality audio files and in reducing the computational cost.
		
	\subsection{DeepFake News} 
		Unlike recency of fake audio and video payloads, the production of fake textual contents such as hoax and rumor has a long history. However, advent of deepfake accelerated fake news production. Eye-catching advancements in natural language processing algorithms resulted in powerful techniques that are capable for generating comparable texts to the human-generated text.  Although a wide range of statistical and ML-based approaches have been used for generating fake texts, GANs-based models are the most dominant.\\ 
		\begin{figure*}[h]
		\includegraphics[width=\textwidth]{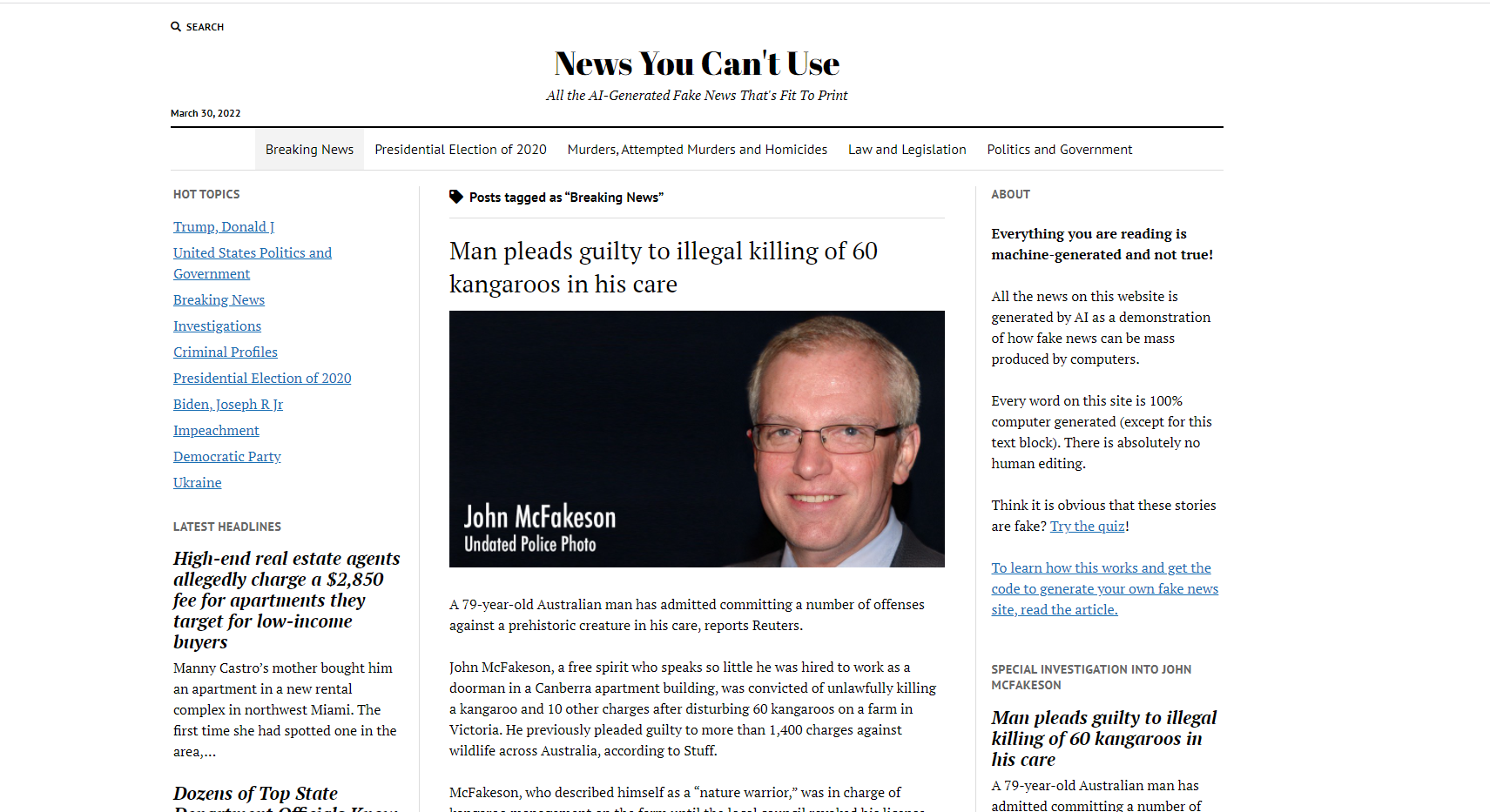}
		\caption{A website where all the news is generated by AI \cite{web6}}
		\label{fig:Fakenew}
	\end{figure*}
		Nie et al.\cite{nie2018relgan} proposed \textit{RelGAN} that included three main components. The first component was a relational memory-based generator to model the long-distance dependency of text contents. Afterwards, a Gumbel-Softmax component was included to train \textit{RelGAN} on discrete data. The last component generated embedded representations. The \textit{RelGAN} achieved performance of over 88\% on BLEU\footnote{In the NLP domain, BLEU is an evaluation metric to compare the machine's generated text with existing human-generated} metric. \textit{Meta-cotga} is another GAN-based approach that introduced a cooperative training paradigm to handle mode collapse through the generator's training period \cite{yin2020meta}. The \textit{Meta-cotga} achieved over 88.2\% of BLEU metric and outperformed \textit{RelGAN}.\\
	    Yu et l.\cite{yu2017seqgan} proposed \textit{SecGAN} which was a sequence generation framework that improved the quality of discrete tokens using reinforcement learning. The \textit{SecGAN}'s BLEU metric was around 56\% in generating fake speech based on Obama's previous talks and 94\% in music generation. In order to improve \textit{SeqGAN} performance, an Objective-Reinforced Generative Adversarial Network (ORGAN) was  developed which considered domain-specific objectives in addition to discriminator rewards and achieved a higher performance in generating domain-specific texts \cite{guimaraes2017objective}.\\
		Guo et al.\cite{guo2018long} introduced \textit{LeakGAN}, a long text generation architecture that leveraged intermediate information during the text generative process to improve the quality of output when the length of generated samples are increased. The \textit{LeakGAN}'s BLEU metric was 95\%  on the COCO dataset in compare with SeqGAN perofrmance of 83\%. Li et al.\cite{pmlr-v89-li19g} presented \textit{JSD-GAN} that altered minimax optimization between discriminative and generative components and utilized Jenson-Shannon Divergence optimization and acheived over 91\% BLEU score over Obama political speech dataset.	
 \begin{table}[htb]
    \small
    \renewcommand\tabularxcolumn[1]{m{#1}}
    \renewcommand\arraystretch{1.1}
	\begin{tabularx}{\linewidth}{|>{\centering}p{4.8em}|>{\fontfamily{qcr}\selectfont}l|>{\centering}p{2.1em}|>{\RaggedRight}X|}
    \Xhline{0.8pt}
	\thead{Category}    & \thead{Reference}  & \thead{Year}     & \thead{Summary}\\
	\hline
		\multirow{16}{=}{Generative  Adversarial   Network}
	    &  Gui et al.\cite{gui2021review}    &    2021       &
		 This survey explains GAN architectures and the relation among different variations of GANs and their evolution process and background. The survey reviews GANs based on their theoretical aspects and mathematical issues as well as applications, including vision, text generation, speech, audio, etc.	    
	    \\ \cline{2-4} 
		& Cai et al.\cite{cai2021gan}        &    2021       &
		Despite the broad utilization of GANs, there are still considerable privacy and security concerns associated with GANs. This survey looks at GANs through a security and privacy lens and reviews different aspects of data and model's privacy. It continues with investigating security aspects, including model robustness as well as cyber security applications such as botnet and malware.		
		\\ \cline{2-4} 
		& Alqahtani et al.\cite{alqahtani2021applications}     &    2021       & 
		This survey starts with a detailed review of GAN's architecture and variants. Then, provides the reader with a comprehensive persuing of GANs' vision and imagery applications. It also investigates GAN's other areas of productivity, such as medicine and biochemistry.
		\\ \cline{2-4} 
		&  Jabbar et al.\cite{jabbar2021survey}    &   2021        & 
		This survey proposes an in-depth review of GANs, their variants, and a detailed investigation of various GANs applications in different research areas. However, the core idea of this survey is to argue the GAN model training issues and their probable solutions.	
		\\ \hline
		\multirow{6}{=}{Fake\\  Image/Video Generation}
		&   aldausari et al.\cite{aldausari2022video}   &     2022      &
		This survey reviews GAN's architecture and different versions. Then, it specifically review and explain GANs designed for synthetic video generation, their related benchmark datasets and evaluation metrics.		
		\\ \cline{2-4} 
		&   Mirsky et al.\cite{mirsky2021creation}   &     2021      &     
		This study provides a detailed review of GANs functionality and then explain the process of creating and detecting fake imagery contents and remarkable trend in this field. The survey also investigates defensive mechanisms' weaknesses and discusses future works. 
		\\ \cline{2-4} 
		& Zhang et al.\cite{zhang2022deepfake}      &     2022      &     
		The survey provides a general review of deepfake generation mainly for face manipulation and then explain detection approaches as well as benchmark datasets.	
		\\ \hline
		\multirow{4}{=}{ Fake\\ Voice\\ Generation}		
		&  Tan et al.\cite{tan2021survey}    &     2021      &         
		Authors comprehensively survey Text-to-Speech(TTS) systems, especially those based on neural networks for generating synthetic voices. They provide an in-depth review of TTS's components. The survey also includes information about benchmark datasets and open-source libraries.		
		\\ \cline{2-4} 
		&   Wenger et al.\cite{wenger2021hello}   &     2021      &    
		This survey presents a thorough experimental study on deep-learning based speech synthesis attacks and demonstrates both human, AI-based detectors and defensive techniques can be fo0led by synthetic voices. 	
		\\ \hline
		\multirow{4}{=} { Fake  News  Generation}
		&  Rosa and Papa\cite{de2021survey}    &     2021      &         
		This study proposes a comprehensive review of recent research on text generation by GANs. The authors analyze and discuss research and introduce benchmark datasets and evaluation metrics.		
		\\ \cline{2-4} 
		& Zhou and Zafarani  \cite{zhou2020survey}    &   2020        &
		Although this survey mainly focuses on fake news detection, it provides detailed information about the theoretical and social aspects of fake news. Then, it proposes a thorough review of fake news detection methods. 		
		 \\ \hline
    \Xhline{0.8pt}
	\end{tabularx}
	\caption{{A curated list of important deepfake generation papers}}
	\label{tab:generation}
 \end{table}

    \section{Deep Fake Detection and Deterrence: State of the Art}\label{sec:deepfake}
 There are increasing evidence of adversaries using deepfake payloads as part of their attack campaigns. Hence, substantial research and development efforts have been made to address this issue. In this section we review important academic and industrial efforts on detecting or deterring deep fake inputs and offer a curated list of references for further studies (see Table \ref{tab:detection}).
	\subsection{Deepfake image/video detection}\label{subsec:detectimage}
 Zhang et al. \cite{8124497} introduced a method based on Speeded Up Robust Features (SURF), Bag of Words (BoW), and Support	Vector Machine (SVM) to detect face swapping in image files. The SURF was responsible for extracting  descriptive features from images to detect any deformation. Afterwards, BoW method embedded extracted features (descriptor) into a more compact space, which was used by the SVM to classify different features to identify fake payloads. They achieved 93\% detection rate over a dataset of swapped faces. Durall et al.\cite{durall2019unmasking} have tried to propose a detection method using the image's signals features. They utilized Discrete Fourier Transform (DFT) to extract features for training different classification algorithms. They trained Logistic Regression, SVM, and K-Means on three datasets of fake images and claimed 100\% detection rate. Agarwal et al., \cite{agarwal2019protecting} proposed a method using classic machine learning for identifying deepfake videos. They first created a dataset of world leaders, celebrities, etc. Then, applied a one-class SVM for novelty detection and considered these novelties as fake contents. They achieved an average Area Under Curve(AUC) of 95\%.\\
 While DL models were used to generate deepfake payloads, many deepfake detection methods are using DL as well \cite{nguyen2019deep}. Mo et al.\cite{mo2018fake} introduced a Deep Convolutional Neural Network (DCNN) architecture that implicitly learned image features and use them to identify fake paylaods. They achieved an average detection rate of 99.3\% outperforming comparable methods. In another DCNN-based study, Shahrooz et al.\cite{shahroz2019} designed an ensemble of shallow benchmark DCNN models called \textit{ShallowNet} and obtained over 99.8\% detection rate. Li and Lyu \cite{li2018exposing} hypothesized that deepfake generators can only generate images/videos with limited resolution. This leads to some identifiable artifacts that can be used to detect fake payloads. They designed a DCNN to learn these artifacts and achieved an average AUC of  98.5\% in detecting deepfake payloads. \\
	Li et al.\cite{8630787} introduced a new fake detection architecture consisting of DCNNs and LSTMs. The DCNN component was responsible for extracting eye blinking features. Afterwards, the LSTM component learned this sequence as a features to discriminate between natural and synthetic eye blinking in videos. Their experiments showed 99\% of AUC. Dang et al.,\cite{dang2020detection} proposed a method that included an attention-based mechanism for preparing features for classification. They coupled DCNN with an attention map layer that generated high-level and distinguishable features for final classification. In another study, Li et al.\cite{li2020face} introduced a new image representation called Face X-ray. They claim that Face X-ray can reveal the face boundary of forged images and achieved an average AUC of 99.1\%. 
	\subsection{Deepfake voice detection}\label{subsec:detectvoice}
	Although deepfake images/videos are constantly in the media spotlight, fake voices can be more dangerous and cause more damages. Deepfake voices can be the main tool of the attack in the criminal campaigns. For instance in 2019, cybercriminals used deepfake voice techniques in a scam campaigns to impersonate the CEO and steal over \$240,000 \cite{albadawy2019detecting}. Furthermore, audible fake payloads can be used as part of spear phishing campaigns to make spam messages more believable. In this section, we review the most important state-of-the-art synthetic voice detectors. \\
	AlBadawy et al.\cite{albadawy2019detecting} presented a detection approach using the bispectral	analysis method. They prepared bispectral artifacts to calculate bicoherence of the given voice file. They statistically analyzed these features and applied a simple classification method on extracted features. Experiments on benchmark voice datasets incited around 99\% of AUC for most synthetic voices.	Ahmed et al.\cite{ahmed2020void} introduced VOID (voice liveness detection) system to distinguish between the spectral power of live-human voices and other voices. The introduced system leveraged a SVM trained with 97 features extracted from the speech spectrum. VOID could successfully detect unreal inputs with an accuracy of 98\%.  \\
	VoiceLive \cite{zhang2016voicelive} was another fake voice payload detection technique with four major components. The first component segmented voices phoneme. The second component calculated time-difference-of-arrival (TDOA) based on the speaker's microphone functionality. The third component explored the similarity of inputted voice with profiles of previous voices stored in a database. Finally, the forth component classified voices based on the calculated similarity score. VoiceLive achieved an average accuracy of more than 99\%. Wang et al.\cite{wang2020deepsonar} used the internal information of Speaker Recognizing (SR) systems for identifying AI-synthesized voices. They employed a DL model from the SR task and monitored its internal parameters (neurons) while processing fake and real speeches. Based on the collected data, they trained a binary-classifier with a shallow neural network and achieved an average AUC of 98\%.
	\subsection{Deepfake news detection}\label{subsec:detectnews}
	Availability of advanced fake text generation tools have yielded to a rapid spread of textual forgeries. Hence, detecting fake text payloads including fake news is imperative.  Several ML-based methods have been proposed for identifying unrealistic texts.  This section reviews the state-of-the-art techniques and reports their performance in fake news detection.\\
	Zubiaga et al.\cite{zubiaga2018discourse} tested performance of different classic machine learning methods such as SVM, Decision Tree, and DL models, including LSTM for detecting Twitter's rumor stance. Their results indicated desirable performance of LSTM (F-measure of 65\%) for fake tweet detection. In a similar approach, Aphiwongsophon and Chongstitvatana\cite{aphiwongsophon2018detecting} employed Naive Bayes, Neural Network and SVM for fake tweet detection where SVM outperformed other classifiers and achieved accuracy of 99.90\%. 
	Reis et al.\cite{reis2019supervised} analyzed importance of different text features to identify fake payloads. They then used the prominent features to evaluate the performance of different supervised ML algorithms for fake news detection and achieved the best performance with XGBoost (F-measure of 81\%). 
	Bahad et al.\cite{bahad2019fake} leveraged recurrent neural network architectures and, by training a bi-directional LSTM  and GloVE word embedding, tried to detect fake news. They also tested DCNN and unidirectional LSTM and demonstrated that bi-directional LSTM is the most accurate detector with 92\% accuracy. 
	Sahoo and Gupta \cite{sahoo2021multiple}, hypothesized that advanced deepfake news can not be detected using textual features only. Hence, they included meta-information such as user-profile, account activities, etc. to train a LSTM and achieved F-measure of 99.3\% outperformed rival models.  In another attempt, Shu et al.\cite{Shu2019profile} utilized a wide range of meta-data such as user profile, user location, etc., to train a LSTM and achieved F-measure average of about 96\%. \\
	In order to take the full advantage of various DL models, Nasir et al.\cite{nasir2021fake} proposed a hybrid approach consisting of both DCNN and LSTM engines. The DCNN module was responsible for extracting local features, while the LSTM  was learning from the long-term dependencies between extracted features. Their results indicated the superior performance of their hybrid model with F-measure of over 99\%. More recently, Kaliyar et al.\cite{kaliyar2021fakebert} utilized BERT transformers, to build a method called FakeBERT to identify fake news with accuracy of over 98.9\%.  
	
 \begin{table}[htb]
    \small
    \renewcommand\tabularxcolumn[1]{m{#1}}
    \renewcommand\arraystretch{1.1}
	\begin{tabularx}{\linewidth}{|>{\centering}p{4.8em}|>{\fontfamily{qcr}\selectfont}l|>{\centering}p{2.1em}|>{\RaggedRight}X|}
    \Xhline{0.8pt}
	\thead{Category}    & \thead{Reference}  & \thead{Year}     & \thead{Summary}\\
	\hline

		\multirow{4}{=}{  Fake\\ Image/Video  Detection}
		& Tolosana et al.\cite{tolosana2020deepfakes}  &     2020      &
		This study comprehensively investigates methods for manipulating deepfake images, primarily for faces. The authors describe various face manipulation trends and their properties. Then,  an up-to-date review of deepfake detections and their functionality is presented.
		\\ \cline{2-4} 
		&  Juefei-Xu et al.\cite{juefei2021countering}  &   2021        &  
     	This survey provides a comprehensive review and in-depth analysis of the research work for deepfake generation and detection as well as evasion of deepfake detection. A helpful taxonomy has been presented that depicts the connection between generators and detectors in previously conducted research. 
		\\ \hline
		\multirow{4}{=}{Fake Voice  Detection}	
		& Balamurali et al.\cite{balamurali2019toward}     &    2019       &   
		For most deepfake detection studies, there is a direct relationship between the quality, importance, and effectiveness of extracted features for the classification phase and detection performance. This survey reviews various features extracted from audios for non-realistic voice detection. 	
		\\ \cline{2-4} 
		& Zhang et al.\cite{zhang2017investigation}  &     2017      &    
		Different DL models, mainly DCNN and LSTMS, are reviewed in this survey, and their application for verifying speakers and identifying fake voices is revealed. 		
		\\ \hline
		\multirow{8}{=}{ Fake  News  Detection}
		&  Sharma et al.\cite{sharma2019combating}  &   2019        &  
		In this review paper, the authors try to describe challenges connected to deepfake news detection technically. Then, state-of-the-art detection and mitigation techniques are explained in-depth.   	
		\\ \cline{2-4} 
		& Mridha et al.\cite{mridha2021comprehensive}   &    2021       & 
		Focusing on DL models, this survey provides the reader with a comprehensive review of DL models for detecting fake news. They covered the gap between past and current literature reviews by including cutting-edge DL models such as Attention mechanisms and Transformers. Also, evaluation metrics for detecting fake news are argued. 	
		\\ \cline{2-4} 
		&  Zhou and Zafarani\cite{zhou2020survey}  &   2020        &	
		Although this survey mainly focuses on fake news detection, it provides detailed information about the theoretical and social aspects of fake news. Then, it proposes a thorough review of fake news detection methods. 		
		\\ \cline{2-4} 
		&  Zhang et al.\cite{zhang2020overview}  &   2020        &	
		The authors highlight the importance and negative impact of online fake news in this survey. Afterward, they provide the reader with a thorough overview of deepfake detection approaches. In addition, they reveal the inclusion of meta-information such as user-profile for many detection methods, while meta-information may fall into the misinformation category				
		\\ \hline
	\end{tabularx}
	\caption{Curated list for deepfake detection studies}
	\label{tab:detection}
\end{table}

\section{Proposed Solution}\label{sec:solution}

\begin{figure*}[h!]
		\includegraphics[width=\textwidth,height=11cm]{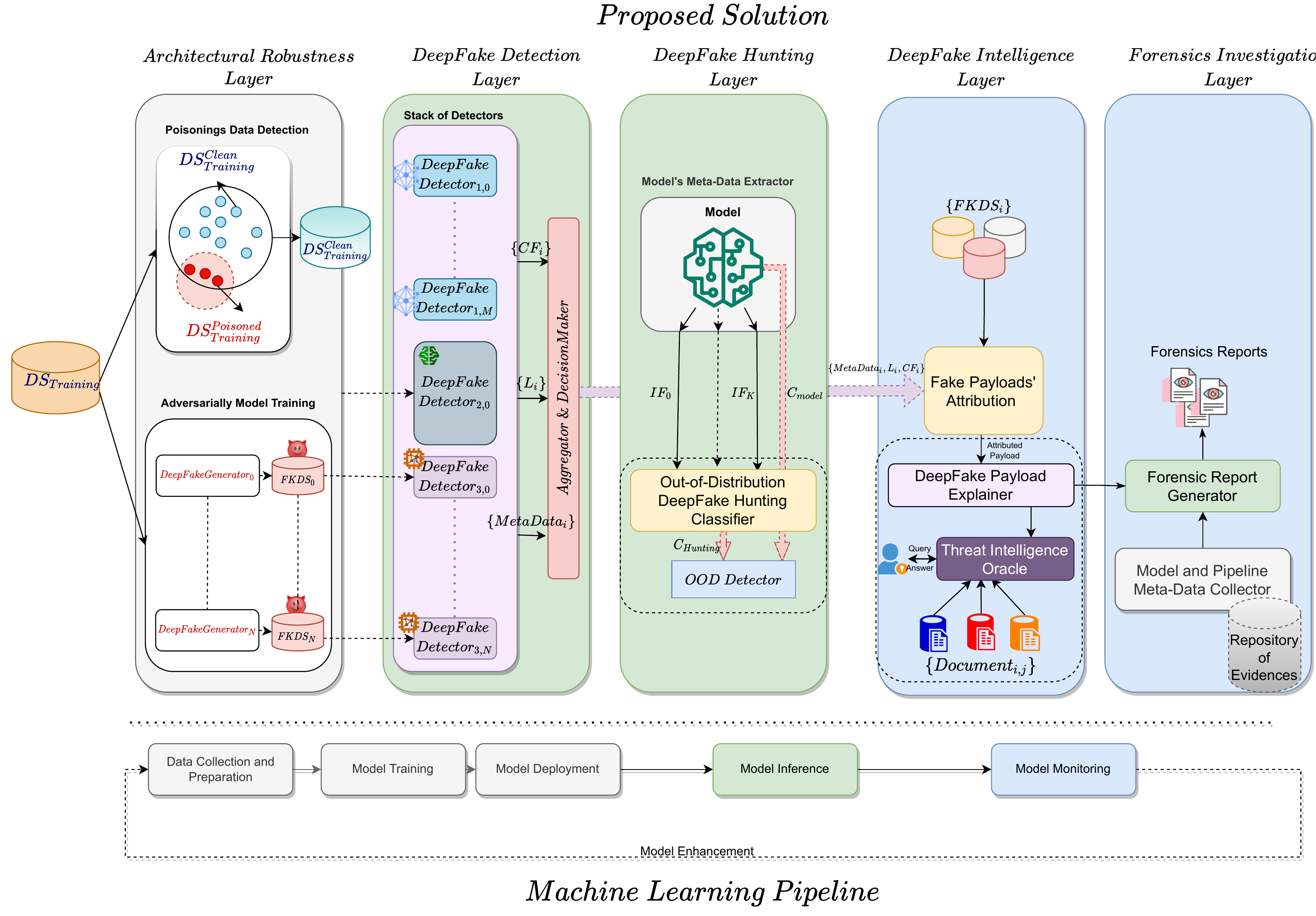}
		\caption{The proposed solution overview}
		\label{fig:solution}
	\end{figure*}

This section offers an overview of our proposed solution and its components and functionalities. Our solution consists of five layers (see Figure \ref{fig:solution}) aligned with SSC layers and the ML pipeline. The first layer elevates the security and robustness of detector models against adversarial examples. The second layer is using a wide range of deepfake detectors to offer a robust deepfake detection service. The third layer actively identifies and hunts unforeseen deepfakes. In order to obtain in-depth knowledge related to sophisticated and state-sponsored deepfakes, the fourth layer is armored by a threat attribution component. Also, this layer is equipped with an intelligent agent for answering complex questions related to identified deepfake payloads. The last layer collects heterogeneous information about the model pipeline and generates forensic reports.
	
\subsection{DeepFake Architectural Robustness Layer}\label{subsec:arcitecturelayer}
	This layer includes two main components to monitor development, training, and testing given ML eco-system and ensures robustness of in-production engines against DeepFake payloads. 
	
	\subsubsection{Poisoning Data Detection Component} \label{subsubsec:l1c1}
\begin{figure*}[h!]
		\centering
		\includegraphics[width=15cm]{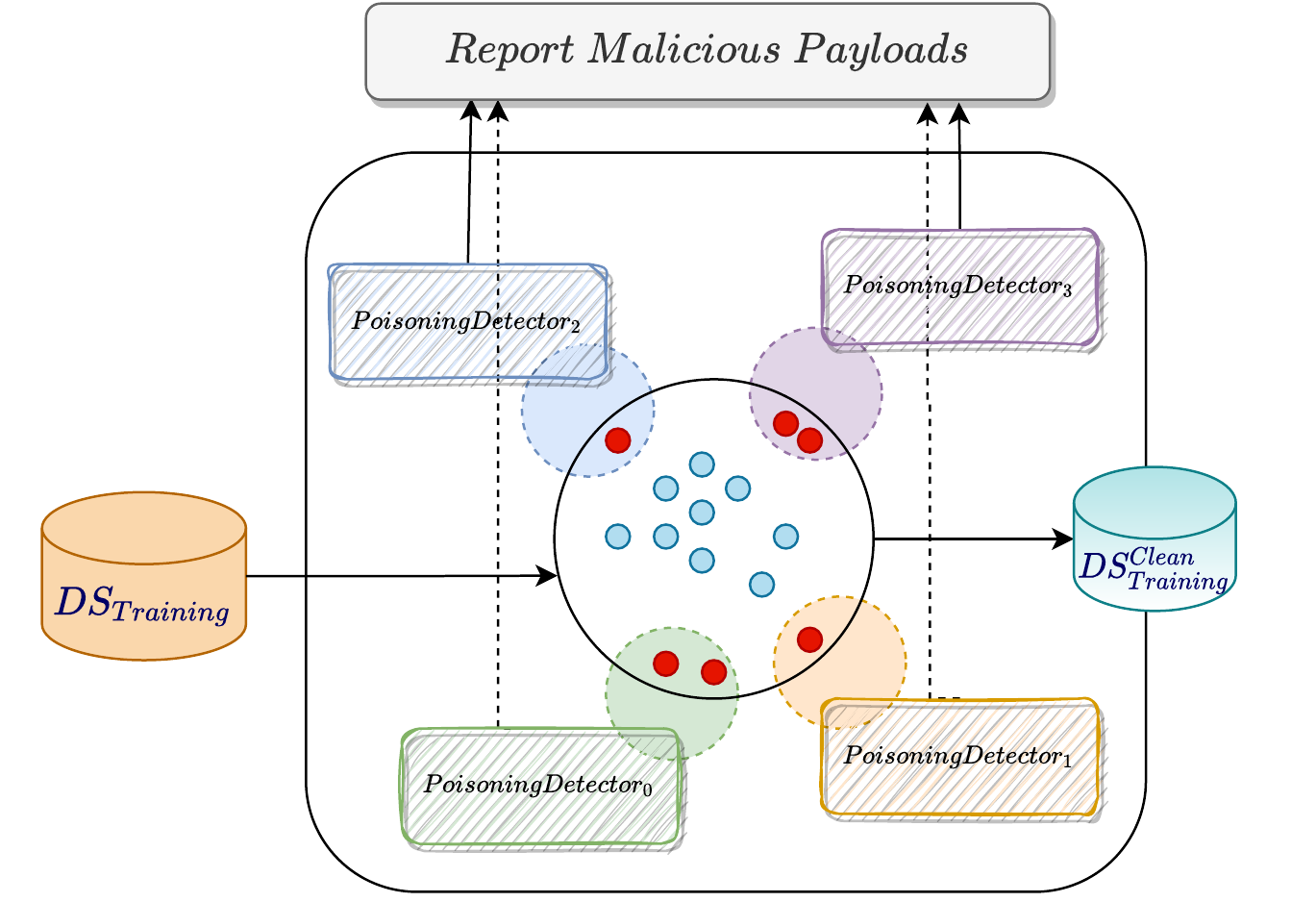}
		\caption{Poisoning Data Detection Component}
		\label{fig:l1c1}
	\end{figure*}

	Poisoning attacks \cite{goldblum2020dataset,schwarzschild2021just} are among sophisticated vulnerabilities of trained ML systems with roots in the model's training data. Attackers try to inject poisoned samples $ DS_{Training}^{Poisoned}$ into the model's training data $ DS_{Training}^{}$ to change the model's decision boundaries during the model's inference time. This component is responsible for exploring model's training data ($ DS_{Training}^{}$) and detecting poisoned samples ($ DS_{Training}^{Poisoned}$) that are injected during training. Moreover, it provides a safe and clean dataset ($ DS_{Training}^{Clean}$) for training tasks.
	Proposed techniques for detecting poisoning inputs look at the data manifold from different lenses according to their hypothetical design and parameters. Biasness toward a detector paves the way for a cybercriminal to poison the training dataset while evading single detector's detection scope. In order to mitigate this issue, the poisoning attack detection component consists of an array of detectors $PoisoningDetector_{i}$ that explore the training data $ DS_{Training}^{}$ and identify poisoning samples or outliers that can degrade the deepfake detection engine performance (see Figure \ref{fig:l1c1}). Finally, the component reports rejected samples for further investigation, model enhancement, and generation of threat intelligence feeds.
		
	\subsubsection{Adversarial Model Training Component}
	\label{subsubsec:l1c2}
\begin{figure*}[h!]
    	\centering
    	\includegraphics[width=15cm]{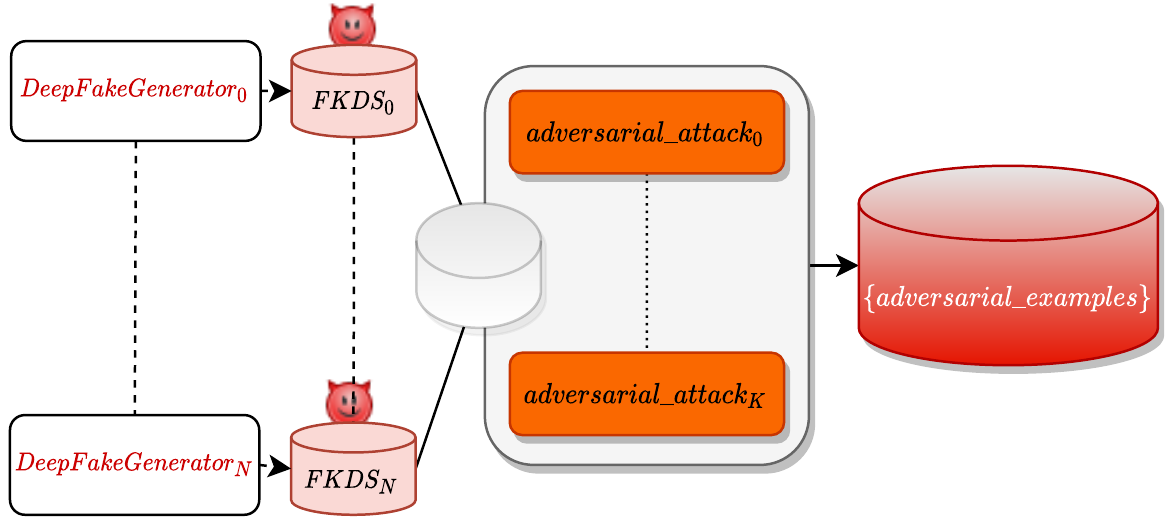}
    	\caption{Adversarially Model Training Component}
    	\label{fig:l1c2}
    \end{figure*}

	Training models with previously known adversarial examples is the main approach to deter adversarial attacks \cite{bai2021recent, silva2020opportunities,tramer2019adversarial}. Adversarial training is a highly recommended solution for enhancing the model's robustness against different security threats including adversarial deepFake payloads \cite{hussain2021adversarial}. As Figure \ref{fig:l1c2} illustrates and at the first stage, this component includes a stack of $N$ generators ( $\{DeepFakeGenerator_{i},i=0,...,N\}$  ). Theses generators are powerful and advanced algorithms devised to generate deepFakes \cite{10.1145/3425780}. Each generator is trained with the detector training dataset $ DS_{Training}^{}$ and outputs a $FKDS_{i}$  $i=0,...,N$ that includes several deepFake samples. The second component prepares a comprehensive dataset of adversarial deepfake payloads ($\{FKDS_{i},i=0,...,N\}$) that are generated by $\{DeepFakeGenerator_{i},i=0,...,N\}$. 
	 State-of-the-art adversarial attacks are employed for generating malicious samples. In the second stage, generated deepfake samples are fed into the adversarial attack array of $\{adversarial\_attack_{i} \ i=0,..., K\}$. These attacks are perturbing $deepfake\_examples\in\{FKDS_{i},i=0,...,N\}$ to generate malicious adversarial examples that are capable of evading detection. Finally, dataset $\{adversarial\_example\}$ is generated for the second round of adversarial training. The training approach could be adopted from  \cite{cheng2020cat,tian2021analysis,alayrac2019labels}. The merged dataset enhances the robustness of the ML model to classify real/fake payloads and to reduce the risk of possible backdoors caused by evasion attacks \cite{hussain2021adversarial}.\\
	This component provides model's stakeholders with a merged dataset including $ DS_{Training}^{Clean}$ and $\{FKDS_{i},i=0,...,N\}$ for re-training the model. Moreover, it filters  out harmful samples and creates $DS_{Training}^{Clean}$ datasets. The main detector model is periodically re-trained using this dataset to increase its robustness and decrease its confidence rate for known adversarial deepfake payloads.\\
	
	\subsection{DeepFake Detection Layer}\label{subsec:detectionlayer}
\begin{figure*}[h]
    	\centering
    	\includegraphics[width=10cm]{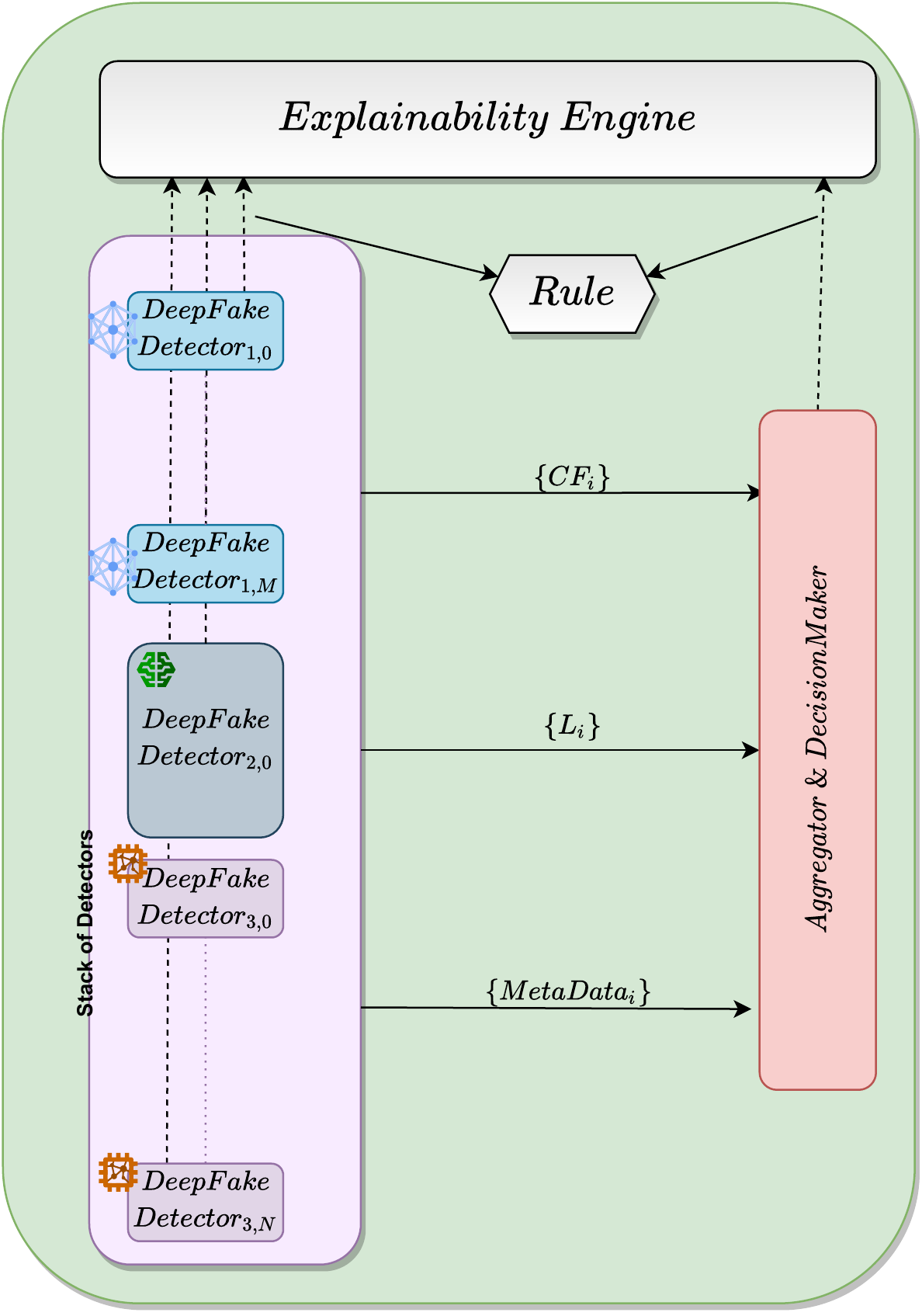}
    	\caption{DeepFake Detection Layer}
    	\label{fig:l2c1}
    \end{figure*}

	The \textit{DeepFake Detection Layer} plays a vital role in detecting - and then rejecting- fake payloads. This layer includes two major components.	The first component is a stack of state-of-the-art DeepFake detectors developed based on the known threat actors' TTPs. The stack benefits from a dynamic structure that allows addition of new detection techniques as patches. Each detector generates a label and a confidence score indicating its detection outcome. The second component receives the output of individual detectors within the stack and aggregates them to make explainable decisions for the given input (see Figure \ref{fig:l2c1}).
    
	\subsubsection{Stack of DeepFake Detectors Component} \label{subsubsec:l2c1}
	This component includes an ensemble of different deepfake detection mechanisms \cite{9506460,hakak2021ensemble,aslam2021fake,bonettini2021video}. During the past decade, several DeepFake detectors have been introduced for identifying unreal inputs to machine learning models \cite{10.1145/3425780, aldwairi2018detecting,9357161,wang2020deepsonar}. However, these detection techniques are based on a specific hypothesis or trained with specific data that transmutes them into local optimum solutions that fail while operating as a holistic approach. To tackle this problem, we empowered the proposed solution with an ensemble mechanism that expands the coverage of the proposed solution. Moreover, this component includes several deepFake detectors that form a stack of DeepFake detectors. The stack accepts the main model's inputs, and each detector generates three main outputs denoted as $L_{i}$ whcih shows the output of $i^{th}$ detector for identifying an input as real or fake. Also, $CF_{i}$ indicates the confidence of $i^{th}$ detector for generating $L_{i}$. Furthermore, each detector provides us with $MD_{i}$ that includes meta-data related to $i^{th}$ detector. $L_{i}$ and $CF_{i}$. $MD_{i}$ are employed for enhancing the explainability of the proposed solution.

	\subsubsection{Aggregation and Explainable Decision Making}\label{subsubsec:l2c2}
	This component receives sets of $L_{i},\ CF_{i}$ and $MD_{i}$ from the detector stack and aggregates detectors' labels - considering detectors' confidence- based on different policies such as majority voting and weighting. This component makes the final decision for accepting or rejecting inputted samples. In addition, this component includes an explainability engine that is attached to each deepfake detector. Then, based on the deepfake detector's structure, the engine extracts several $Rule$s from detector \cite{burkart2021survey}. In a similar manner, rules are extracted from decision making and aggregation module. Finally, the component generates a report that illustrates how the decision has been made based on the stack's information. Also, the report includes extracted knowledge from $\{MD_{i}\}$ that provides decision makers with information about malicious properties that led to detecting a payload.

	\subsection{DeepFake Hunting Layer}\label{subsec:huntinglayer}
	Despite significant endeavors for enhancing intelligent agents' robustness against unknown and malicious inputs, they are still generating highly confident outputs for unknown samples. Generally, a model is trained with an ID dataset and it is expected that the model faces inputs that are statistically and contextually similar to the samples in the ID dataset. However, during operation the model may receive various inputs that are not drawn from ID (OOD samples). 
	Surprisingly, the model may generate highly confident labels for OOD inputs \cite{DBLP:journals/corr/abs-2110-11334, DBLP:journals/corr/abs-2108-13624,  pang2021deep, salehi2021unified}, hence we need to include a hunting layer that identifies OOD examples. 
	To this end, we included a component to extract the model's pattern while processing inputted samples. Then, another component is learning the internal pattern(s) of the model for ID samples to discriminate between ID and OOD samples.
	\subsubsection{Model's Meta-Data Extractor Component}\label{subsubsec:l3c1}
	For identifying unforeseen inputs that cause high confidence prediction - and classifying them as real payloads - our solution is armed with an OOD sample detection component which requires the model meta-data. This component monitors the model and extracts its internal states to identify meta-data required fordDeepfake payloads hunting. This component encapsulates the model and extracts its internal states $\{IF_{i},i=0,...,K\}$ as well as model's output $C_{model}$. For instance, it extracts the given deep learning model's neurons' values or internal parameters. The extraction phase take place at two different stages. The first stage includes extracting $IF_{i}$s for ID samples and the second stage is for meta-data extraction during model's inference stage.
	\subsubsection{Out-of-Distribution Deepfake Hunting Component}\label{subsubsec:l3c2}
	This component is a traditional deep learning classifier trained with internal pattern of activities ($IF_{i}$) extracted from the model while feeding ID samples. At this phase, the component uses  $\{IF{i}\}$ from the main model while processing ID inputs \cite{mohseni2020self,vyas2018out,fort2021exploring}.\\
	During the model's inference time, the component receives model's output(s) as well as meta-data and passes these information to its internally trained model to generate a label ($C_{Hunting}$). To identify a samples as OOD or ID, the component uses following reasoning:\\
	\begin{equation}
		\left\{ \begin{array}{cl}
			if\ C_{model}=C_{Hunting} & \rightarrow  accept \\
			else\ \ \ \ \ \ \ \ \ \ \ \ \ & \rightarrow  {\color{red}\textbf{reject}}
		\end{array} \right.
		\label{eq:hunying}
	\end{equation}
	
	Ovreall this component rejects DeepFake inputs that 1) pretend as real samples and 2) the main model assigns a highly confident score to them during classification while their model's activity pattern is inconsistent compared with real samples.
	
	\subsection{DeepFake Intelligence Layer}\label{subsec:intelligencelayer}
	Identifying the source of attacks and attackers TTPs plays an essential role in designing the best defensive solutions. This layer is responsible for identifying the source of attacks based on the deepfake payloads' properties and assist the model's stakeholders in obtaining deeper insights into the identified threat. This layer consists of two main components. The first component is responsible for attributing fake inputs to their original generation technique(s), and the second is a threat intelligence oracle.
	
	\subsubsection{Deepfake Attribution Component} \label{subsubsec:l4c1}
	This component attributes detected fake payloads to their original actors, generators, or any known campaigns. To this end, the component receives $\{MD_{i},{L_{i},CF_{i}}\}$ from the \textit{Detection Layer}.  Also, using advanced similarity measurements, the component compares detected fake inputs with a huge repository of previously know fake contents. Finally, it assigns the most likely label to the fake input that indicates the attack source. 
	
	\subsubsection{Deepfake Threat Intelligence Oracle Component} \label{subsubsec:l4c2}
	
	This component consists of two main parts. The first part is a deepfake explainer module that explains the properties of fake payloads that resulted to a detection or attribution. This part can be used by other components that require exaplainability (see Section \ref{subsec:forensicslayer}). 
	The second part is an advanced Natural Language Processing (NLP) agent\cite{devlin2018bert} that is trained with heterogeneous documents related to fake payload generators ($\{Document_{i,j}, i=0,...,NumberOfAttackers\}$). The oracle receives explainable information related to attributed payloads as well as a security specialist's queries. Then, the oracle provides comprehensive answers that include insights into different aspects of the attack. These queries can be defined as fixed questions, and the oracle outputs can be formatted as advanced threat intelligence reports.
	\subsection{DeepFake Forensics Layer}\label{subsec:forensicslayer}
	Presenting evidences related to detected, rejected and attributed fake payloads is an essential property of the proposed solution.  This layer is responsible to generate forensic reports and contains two components 
	\subsubsection{Model and Pipeline Meta-Data Extractor Component}\label{subsubsec:l5c1}
	In order to prepare a comprehensive and insightful forensics report, it is necessary to collect a wide range of evidences and then build possible scenarios based on those evidences. This component is responsible to assess the model's pipeline and collect various evidences and stores them in its repository \cite{mena2016machine}.
	\subsubsection{Forensic Report Generator Component}\label{subsubsec:l5c1}
	This component includes an advanced querying tool that deeply explores collected pieces of evidence stored in a repository and relates them to each other. Afterward, the component parses explored shreds of evidence and generates a forensics report that include explainable data generated in Section \ref{subsec:intelligencelayer}.

	\section{Assessment and Analysis Metrics} \label{sec:criteria}
	In cybersecurity, it is vital to always monitor and measure the performance of different solutions that are deployed in an enterprise and deepfake security solutions are not an exception. This section  provides a clear view of possible metrics and measurements that can be used to evaluate and monitor performance of the our proposed solution at different layers of SSC. 
	
	\begin{itemize}
		\item \textit{DeepFake Architectural Robustness Layer}: This layer consists of two main components. The Poisoning Data Detection component can be evaluated based on its detection rate that reflects the ratio between detected versus all poisoned samples. The second component (Adversarially Model Training) enhances the model's robustness by including adversarial samples. Therefore, various robustness metrics including \textit{Attack success rate, Distortion, CLEVER score} and \textit{Transferability} can be used to monitor its performance \cite{su2018robustness}.
		\begin{itemize}
		\item \textbf{Detection Rate}: In order to evaluate performance of proposed solution in detecting poisoning attacks, a set of $n$ adversarial examples are generated for each algorithm $\{adv\textunderscore example_{alg_{j},i}, i=0,....,n \ \& \  alg_j\in\{poisoning \ attack \ algorithms\}	\}$ . The component detection rate against each algorithm can be calculated as follow:\\
		\begin{equation}
		DetectionRate_j=\frac{|Detected \ adversarial \ example(s)|}{n} \ \ \ \ \ \ \  j\in[1,|attack \ algorithms|]
		 \label{eq:detectionrate}
		\end{equation}
		\item \textbf{Attack Success Rate}: This metric indicates the attacks success rate in terms of bypassing a DL model. After each adversarially training cycle. Using state-of-the-art adversarial evasion attack algorithms, a set of $m$ adversarial examples are generated for each algorithm $\{adv\textunderscore example_{alg_{j},i}, i=0,....,m \ \& \  alg_j\in\{evasion \ attack \ algorithms\}\}$ . The success rate for each attack algorithm $j$ is calculated as follow:\\  
				\begin{equation}
		SuccessRate_j=\frac{|Successful \ adversarial \ example(s)|}{m} \ \ \ \ \ \ \  j\in[1,|attack \ algorithms|]
		 \label{eq:successrate}
		\end{equation}
	    \item \textbf{Distortion:} This metric evaluates the distance  (perturbation) between an adversarial example and its original benign sample. Obviously, higher Distortion indicates higher robustness. In order to calculate this performance metric, a set of $p$ adversarial examples using different algorithms are generated $\{adv\textunderscore example_{alg_{j},i}, i=0,....,p \ \& \  alg_j\in\{evasion \ attack \ algorithms\}\}$. The Distortion metric for each algorithm is calculated as follow:
	    \begin{equation}
		Distortion_j=\frac{\sum_{i=0}^{p}\left \| successful\textunderscore adv \textunderscore examplei,j- original \textunderscore example_{advexamplei,j} \right \|_\infty }{p} 
		 \label{eq:distortion}
		\end{equation}
	    \item \textbf{CLEVER:} Wang et al.\cite{weng2018evaluating} presented a robustness evaluation metric using extreme value theory. CLEVER (Cross-Lipschitz Extreme Value for nEtwork Robustness) is an attack-agnostic approach and calculates the lower bound of adversarial distortion.  For each input $x_0$, CLEVER finds the minimum distortion $\delta$ which is required to misclassify the distorted (adversarial) example $x_0+\delta$ by the detector model \cite{weng2018evaluating}. 
	    \item \textbf{Transferability:} In order to measure this metric and using each attack algorithm $alg_j\in\{evasion \ attack \ algorithms\}$, a set of adversarial examples is generated on the source model $m_{source}$ $\{adv\textunderscore example_{alg_{j},i,source}, i=0,....,p\}$. Then, \textit{Attack Success Rate} of these examples on the target model $m_{target}$ is measured that indicates the transferability of adversarial examples from $m_{source}$ to $m_{target}$. The lower Transferability means higher robustness.
		\end{itemize}
		
		\item \textit{DeepFake Detection Layer}: The proposed \textit{Stack of DeepFake Detectors} is the main component in this layer. The stack includes several deepfake detectors, and each detector can be evaluated using traditional AI metrics such as A\textit{ccuracy, AUC, F-measure, Precision} and \textit{Recall} . Furthermore, explainability of the model can be evaluated using interpretability metrics such as  $\{D,\ R,\ F,\ S\}$. 
		\begin{itemize}
		\item \textbf{Accuracy\footnote{  
		True Negative (TN) is the number of real examples identified as real.
		True Positive (TP) is the number of Positive examples identified as Positive.
		False Negative (FN) is the number of deepfake examples identified as real.
		False Positive (FP) is the number of real examples identified as deepfake.
		}:} Accuracy is the widely reported for classification tasks and is calculated as follow:
    		\begin{equation}
                \label{eqn:acc}
                Accuracy = \frac{TP + TN}{TP + FN + TN + FP}
            \end{equation}
		\item \textbf{Precision:} This metric measures the ratio of precise positive (deepfake) prediction over all positive prediction as below:
		\begin{equation}
            \label{eqn:precision}
            Precision = \frac{TP}{TP + FP}
        \end{equation}
		\item \textbf{Recall:} This metric indicates the correct positive prediction over the value of a specific class as:
		\begin{equation}
            \label{eqn:recall}
            Recall = \frac{TP}{TP + FN}
        \end{equation}
		\item \textbf{F-measure:} This indicator calculates the weighted average of the precision and recall as below:
		\begin{equation}
            \label{eqn:F-measure}
            F1-score = 2 * \frac{Precision * Recall}{Precision + Recall}
        \end{equation}
		\item \textbf{D.R.F.S:} where \textbf{D},  calculates the difference between model's performance and the logic of explanation; \textbf{R}, counts the number of rules in the explanation; \textbf{F}, counts the number of features to build explanation; and \textbf{S}, measures the stability of explanation \cite{rosenfeld2021better}.
		\end{itemize}
		
	\item \textit{DeepFake Hunting Layer}: Identifying OOD inputs that the previous layers could not detect is the primary responsibility of this layer. Therefore, a set of metrics applicable on OOD and novelty detection, including \textit{AUC, False Positive Rate at 95\% True Positive Rate, Area under the Precision-Recall curve, Accuracy, and F-measure } can be used to measure performance at this layer \cite{salehi2021unified}.
	\item \textit{DeepFake Intelligence Layer}: This layer consists of two components. The first component attributes a malicious fake payload to its generator. Therefore, general classification performance metrics such as $Accuray,\ AUC,\ F-measure,\ Precision,\ Recall $ can be used for evaluation. Also, the layer includes an intelligent oracle agent that replies to users' queries. Hence, question-answering and dialogue systems evaluation metrics such as BLEU, METEOR, and ROUGE can be used to assess the performance of question-answering task at this layer\cite{liu2016not}. BLEU \cite{chen2019evaluating} is a precision-based indicator that is developed to compare machine generated translation texts with the reference text. On the other hand,  METEOR and ROUGE are F-measure based metrics for identifying performance of a question answering agent for translation and summarization\cite{chen2019evaluating} . For the proposed solution, a corpus of reference question-answering textual data as well as the layer's outputs will be compared using these evaluation metrics.
	\item \textit{DeepFake Forensics Layer}: The primary role of this layer is to generate various forensic reports based on user requests. These reports include all previously calculated metrics as well as information retrieved by the layer's \textit{Model and Pipeline Meta-Data Extractor} component. Most of this layer's reports can be assessed by a human expert. 
	\end{itemize}

	\section{Conclusion and Recommendations}\label{sec:conclusion}
    In the past few years, AI technologies boast their capabilities in generating fake synthetic payloads resembling realistic human-generated content. These Deepfakes are becoming a severe concern for cybersecurity experts, politicians, and sociologists. Deepfakes can be used as the means to deliver weaponized and malicious payloads and quickly become the core part of a cyber warfare strategies. Hence, there is a growing effort on developing tools, tactics, and techniques to detect and deter deepfakes. In spite of extensive efforts for detecting deefakes even utilizing AI for countering AI, the alarming increase of deepfakes portends cybercriminals' superiority. This study reviewed major trends for misusing AI, specifically deep learning, for generating visionary, audible, and textual forgeries. This study is a strong indication of an arms race between deepfake generators and detectors. According to the identified shortages and gaps, a holistic solution was proposed to detect and deter deepfakes and to mitigate their adverse impacts. The proposed solution consists of five fundamental layers for enhancing robustness of ML systems, improving deepfake detection, building deepfake hunting capacities, providing in-depth deepfake intelligence, and preparing thorough forensic reports. Based on the proposed solution, following recommendations can be made.
 
	\begin{itemize}
		\item Unlike adversarial examples that are wisely manipulated inputs to bypass an AI agent, deepfakes are contents generated by DL to fool humans. Although researchers have tried to detect deepfakes using AI, susceptibility of AI agents to adversarial payloads made them vulnerable to attacks generating deepfake adversarial payloads.  Hence, any efforts to leverage AI for deepfake detection should take into account adversarial robustness as well! 
		\item Our detailed literature review demonstrated that existing deepfake detectors are negatively biased toward specific known features of fake payloads. For instance, focusing on eye blinks, exploring specific unusual features from a speech, or inconsistency in a text. However, it is trivial for an adversary to partially retouch these features and evade detection. Hence, it is important to build a multi-view system that consists of several local detection mechanisms each of which validates different micro-features.  The system can also consider macro-features at a higher level. For instance, it can decide based on the inconsistency among speech content, picture, and voice of a public speaker all at the same time. Furthermore, explainability of the decisions made by the detector is a crucial requirement for forensic soundness of the system - something that should be considered during the design stage of the system.
		\item Recently, it is revealed that DL models may generate highly confident outputs for inputs which are not previously seen by the model (OOD samples). For instance, a deepfake detector may assign a highly confident "real" label to a new (unforeseen) category of deepfakes which undermines its trustworthiness. Hence, it is indispensable to always evaluate robustness of deepfake detectors against OOD examples and hunt for potentially malicious OOD contents. 
		\item Deepfake payloads can be part of an advanced state-sponsored attack campaign. Detecting these contents without knowledge of attacker's TTPs could be very time and resource consuming. Therefore, it is important to develop a system for attributing deefakes to potential campaigns. Furthermore, while an agile investigation and decision-making are imperative for security experts during incident triage, it is challenging to review an enormous deepfake repository in a timely manner. Hence, empowering the system with an advanced agent that plays the role of a cyber-oracle will significantly reduce response time to incidents involving deepfakes. 
		\item Although detecting, hunting, and attributing deepfakes are fundamental deterrence steps, being able to forensically examine collected deepfakes could be even more essential for evaluating their risks.  Hence, we need to build systems to collect, preserve and analyse evidence related to deepfake payloads and prepare professional forensic reports - something that is not attempted so far. 
	\end{itemize}


\bibliographystyle{IEEEtran}  
\bibliography{references.bib}

\end{document}